\documentclass{elsart}
\usepackage{latexsym}
\usepackage{graphicx}
\usepackage{amssymb}
\usepackage{epsfig}
\usepackage{times}
\usepackage{pstricks}
\usepackage{pst-grad}
\usepackage{mathptm}
\usepackage{subfigure}
\usepackage{wrapfig}
\usepackage[english]{babel}
\usepackage{graphicx} 
\newcommand{\sq}[0]{$^{2}$\ }

\fussy

\begin{document}

%------------------------------------------------------------------------------
\begin{frontmatter}

\title{The Cosmic--Ray Proton~and~Helium Spectra measured with~the~CAPRICE98~balloon experiment}

\author{M. Boezio, V. Bonvicini, P. Schiavon, A. Vacchi, and N. Zampa}
\address{University of Trieste and Sezione INFN di Trieste, Via A. Valerio 2, I--34147 Trieste, Italy}

\author{ D. Bergstr\"om}, \author{ P. Carlson}, \author{ T. Francke}, \author{P. Hansen}, and \author{E. Mocchiutti\corauthref{em}} 
\corauth[em]{Corresponding author. Tel.: +46--8--55378186, Fax: +46--8--55378216.}
\ead[email]{mocchiut@particle.kth.se}
\address{Royal Institute of Technology (KTH), AlbaNova University Center (SCFAB),\\ S--10691 Stockholm, Sweden}

\author{M. Suffert}
\address{Centre des Recherches Nucl\'eaires, BP20, F--67037 Strasbourg--Cedex, France}

\author{M. Hof, J. Kremer, W. Menn, and M. Simon}
\address{Universit\"at Siegen, 57068 Siegen, Germany}

\author{M. Ambriola, R. Bellotti, F. Cafagna, F. Ciacio, M. Circella}, \author{ and C. N. De Marzo}
\address{University of Bari and Sezione INFN di Bari, Via Amendola 173,\\ I--70126 Bari, Italy}

\author{N. Finetti\thanksref{aquila}}, \author{ P. Papini}, \author{ S. Piccardi}, \author{ P. Spillantini}, \author{ and E. Vannuccini}
\address{University of Firenze and Sezione INFN  di Firenze,\\ Largo E. Fermi 2, I--50125 Firenze, Italy}

\author{S. Bartalucci, and M. Ricci}
\address{INFN -- Laboratori Nazionali di Frascati, Via E. Fermi 40,\\ CP 13, I--00044 Frascati, Italy}

\author{M. Casolino, M. P. De Pascale, A. Morselli, P. Picozza}, \author{ and R. Sparvoli}
\address{University of Roma ``Tor Vergata'' and Sezione INFN di Roma II,\\ Via della Ricerca Scientifica 1, I--00133 Roma, Italy}

\author{J. W. Mitchell, J. F. Ormes, S. A. Stephens}, \author{ and R. E. Streitmatter}
\address{Code 661, NASA/Goddard Space Flight Center, Greenbelt, MD 20771, USA}

\author{U. Bravar, and S. J. Stochaj}
\address{R. L. Golden Particle Astrophysics Lab, Box 3--PAL, New Mexico State University,\\ Las Cruces, NM 88003, USA}

\thanks[aquila]{Now at Dipartimento di Fisica dell'Universit\`a dell'Aquila, L'Aquila, Italy.}

%##############################################################################
\begin{abstract}
%##############################################################################
A new measurement of the primary cosmic--ray proton and helium fluxes from 3 to 350~GeV was carried out by the balloon--borne CAPRICE experiment in 1998. This experimental setup combines different detector techniques and has excellent particle discrimination capabilities allowing clear particle identification. Our experiment has the capability to determine accurately detector selection efficiencies and systematic errors associated with them. Furthermore, it can check for the first time the energy determined by the magnet spectrometer by using the Cherenkov angle measured by the RICH detector well above 20~GeV/n. The analysis of the primary proton and helium components is described here and the results are compared with other recent measurements using other magnet spectrometers. The observed energy spectra at the top of the atmosphere can be represented by (1.27$\pm$0.09)~$\times$~10$^{4}$~E~$^{-2.75\pm0.02}$~particles~(m$^2$~GeV~sr~s)$^{-1}$, where E is the kinetic energy, for protons between 20 and 350~GeV and (4.8$\pm$0.8)~$\times$~10$^{2}$~E~$^{-2.67\pm0.06}$~particles~(m$^2$~GeV~nucleon$^{-1}$~sr~s)$^{-1}$, where E is the kinetic energy per nucleon, for helium nuclei between 15 and 150~GeV~nucleon$^{-1}$. 

\end{abstract}


\end{frontmatter}
%------------------------------------------------------------------------------

\newpage

%##############################################################################
\section{Introduction}
%##############################################################################

Accurate measurements of the spectra of primary cosmic rays have deep astrophysical implications, since they provide important information on the mechanisms of production of cosmic rays and of the matter distribution in the interstellar space. In fact, the spectral shapes of proton and helium nuclei fluxes are sensitive indicators of the processes of particle acceleration, and the observed fluxes are the primary measure of the energy density of cosmic rays in the interstellar medium. Their spectra also serve as important inputs to calculations that aim to predict the secondary antiproton or positron spectra, which result from high--energy interactions of protons and helium nuclei with the interstellar gas.

Recently, the importance of the normalization of the primary cosmic--ray flux has been emphasized in connection to the atmospheric neutrino observations performed by underground experiments, e.g. Super--Kamiokande \cite{fukuda}. A correct interpretation of these measurements depends on the accuracy of the predictions to which they are compared. The assumptions about the flux of cosmic rays that impinge on the Earth turn out to be among the main sources of inaccuracies in  the simulation of atmospheric showers \cite{gaisser01}. For energies above 10~GeV, where the effect of the solar modulation is less than about 10\%, there is still a difference of about 10\%--20\% in the absolute fluxes published in the recent years \cite{boezio99b,sanuki,alcaraz,seo,bellotti,menn}. 

From this experimental scenario, the need arises for measurements extended over a large energy range and with a good understanding of the systematic uncertainties associated with them. Measurements of primary particles have been carried out using different techniques: magnet spectrometers, e.g. \cite{seo,bellotti}, and RICH detectors \cite{buckley,diehl} have been used for energies up to 100--200 GeV/n, while calorimetric measurements extend to higher energies, e.g. \cite{silverberg,asakimori}. However, comparisons among such measurements show sometimes significant discrepancies.

The differences are probably due to imprecise or incomplete knowledge of detectors response functions during the flight. Monte~Carlo simulations or calibrations at low energy in the laboratory prior to flight are not ideal since experimental conditions in the gondola such as temperature or pressure may not be stable during the flight. If the response function is based on the performance of individual detectors in the laboratory, systematic uncertainties are likely when operated with other detectors and would introduce a bias. So, the best experimental determination of the efficiencies and calibrations is to make use of redundant detectors to select a set of ``good'' events independent of the other detectors. These sets of events can be used to determine the response function of the detectors which are not involved in the selection.

In this paper we report a new measurement of the primary proton and helium nuclei spectra with the CAPRICE98 instrument, which combines different detector techniques. This instrument, described in section \ref{c98}, consisted of a superconducting magnet spectrometer, a Ring--Imaging Cherenkov detector and an imaging calorimeter. With these independent detectors, it was possible to accurately determine both the efficiency and the systematic error associated with each detector. The data analysis is described in section \ref{datana}, which includes the determination of detector efficiencies, systematic errors and the corrections applied. The final results and discussions are presented in section \ref{finres}. The results cover a range of kinetic energy from 3 to 350~GeV for protons, and from 0.9 to 150~GeV~nucleon$^{-1}$ for helium nuclei. 

%##############################################################################
\section{The CAPRICE98 experiment}
\label{c98}
%##############################################################################

The balloon--borne CAPRICE (Cosmic AntiParticle Ring Imaging Cherenkov Experiment) instrument was flown from Ft.\ Sumner, New Mexico, USA (34.5$^{\circ}$ north latitude, 104.2$^{\circ}$ west longitude) to Heber, Arizona, USA (34.3$^{\circ}$ north latitude, 111.0$^{\circ}$ west longitude) on 1998 May 28 and 29 \cite{ml} at a vertical rigidity cut--off of about 4.3~GV \cite{shea}. The data analyzed for this work were collected at an average atmospheric depth of about 5.5 g/cm$^2$ (atmospheric pressure of 4.0~to~5.1~mbar, altitude of 36.0--38.2~km), during an exposure time of almost 21 hours. 

\begin{figure}[!hbtp]
\vspace*{0.5cm}
\includegraphics[width=1.\textwidth]{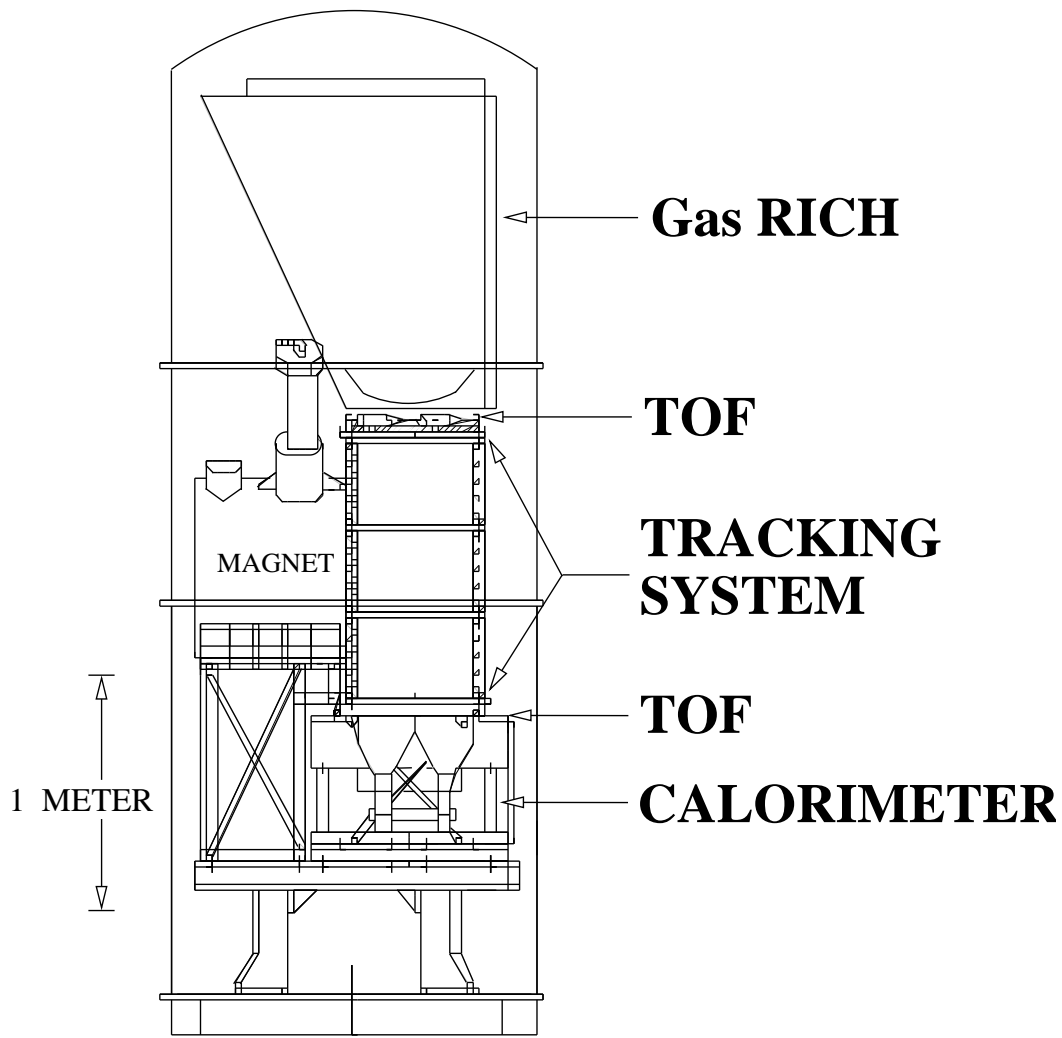}
\vspace*{0.5cm}
\caption{The CAPRICE apparatus in the 1998 configuration (CAPRICE98).
}
\label{caprice}
\end{figure}

The experimental setup was a renewed configuration of the CAPRICE94 apparatus \cite{boezio99b} which was successfully used in a previous balloon experiment at low geomagnetic cut--off in 1994. The CAPRICE98 apparatus \cite{ml} is shown in Fig. \ref{caprice}: it consisted of, from the top to the bottom, a gas Ring Imaging CHerenkov detector (Gas RICH), a time--of--flight device (TOF), a superconducting magnet spectrometer (Tracking system), and a silicon--tungsten imaging calorimeter. 

%&&&&&&&&&&&&&&&&&&&&&&&&&&&&&&&&&&&&&&&&&&&&&&&&&&&&&&&&&&&&&&&&&&&&&&&&&&&&&&
\subsection{The Gas RICH detector}
%&&&&&&&&&&&&&&&&&&&&&&&&&&&&&&&&&&&&&&&&&&&&&&&&&&&&&&&&&&&&&&&&&&&&&&&&&&&&&&

The RICH detector \cite{francke,david,davidl} used a 1 m tall gas (C$_4$F$_{10}$, $\gamma_{th}\simeq19$) radiator and a photosensitive multi--wire proportional chamber (MWPC) mounted above it. The Cherenkov photons were reflected into the chamber by a spherical mirror located at the bottom of the radiator. The chamber was filled with TMAE\footnote{Tetrakis--dimethyl--amino--ethylene, a photosensitive gas; the Cherenkov light interacts with TMAE producing photoelectrons.}--saturated ethane gas. The signals  were acquired  by means of a pad readout implemented on one cathode plane. This plane had an area of 51.2$\times$51.2~cm$^{2}$, divided in 64$\times$64 pads of size 8$\times$8~mm$^{2}$, where the cone of Cherenkov light gave a ring--like image. The ring diameter, dependent on the velocity of the particle, increased from 0 at the RICH threshold (about 18~GV for protons) to about 11~cm ($\sim$50~mrad Cherenkov angle) for a $\beta\simeq$1 particle. About half of the particles triggered by the instrument passed through the MWPC, where they ionized the gas. The ionization signals were amplified and detected by the pad plane along with the Cherenkov signals.

A single photoelectron detected by the multi--wire proportional chamber was collected by 3--5 pads. For $\beta\simeq$1 singly charged particles, an average of 12 photoelectrons per event were detected.  

This RICH detector was designed primarily to identify antiprotons in the cosmic rays against a large background of electrons, muons and pions \cite{carlson}.

%&&&&&&&&&&&&&&&&&&&&&&&&&&&&&&&&&&&&&&&&&&&&&&&&&&&&&&&&&&&&&&&&&&&&&&&&&&&&&&
\subsection{The time--of--flight system}
%&&&&&&&&&&&&&&&&&&&&&&&&&&&&&&&&&&&&&&&&&&&&&&&&&&&&&&&&&&&&&&&&&&&&&&&&&&&&&&
The time--of--flight (TOF) system consisted of two planes of plastic scintillators, located immediately above and below the tracking stack. Each plane was segmented into two paddles viewed at opposite ends by 5~cm diameter photomultiplier tubes. Each paddle had a size of 25$\times$50~cm$^{2}$ and a thickness of 1 cm. The signals from each photomultiplier  were independently digitized for time--of--flight measurements as well as for pulse height analysis. The scintillator signals also provided the trigger for the data acquisition system.

The distance between the two scintillator layers was 1.2~m and the system had a time resolution of about 230 ps.

%&&&&&&&&&&&&&&&&&&&&&&&&&&&&&&&&&&&&&&&&&&&&&&&&&&&&&&&&&&&&&&&&&&&&&&&&&&&&&&
\subsection{The tracking system}
%&&&&&&&&&&&&&&&&&&&&&&&&&&&&&&&&&&&&&&&&&&&&&&&&&&&&&&&&&&&&&&&&&&&&&&&&&&&&&&
The magnet spectrometer consisted of a single coil superconducting magnet \cite{golden78}, which had been used in all the previous flights operated by the WiZard Collaboration \footnote{The WiZard Collaboration, with members from France, Germany, Italy, Sweden and USA, is involved in a long--term investigation of primary cosmic rays with balloon--borne and satellite detectors.}, and a tracking device consisting of three modules of drift chambers \cite{hof94}. The total height of the spectrometer was about 110 cm. 

The lateral sides of each chamber box were made from 1~cm thick epoxy--composite plates, while the open top and bottom sides were covered with 160~$\mu$m thick copper plated mylar windows. The inner gas volume of each box was of size \linebreak 47$\times$47$\times$35~cm$^{3}$. The drift chamber had six layers, each containing sixteen \linebreak 27.02~mm wide drift cells, for measurements in the x--directions and four layers for the y--direction. This system performed 18 measurements along the direction of maximum bending ($x$) and 12 measurements along the perpendicular view ($y$), with a resolution better than 100~$\mu$m. 

The alignment of each drift chamber with the whole tracking system is important for a precise rigidity determination from the measured deflection. This was done by comparing the extrapolation of the track fitted by one chamber to the track fitted by all the drift chambers. Any statistically significant differences in the fitted tracks would then be due to misalignment between the drift chambers. Thus, the position of each wire in one drift chamber was determined precisely by comparing the data from the fitted tracks using all the drift chambers with the position provided by the data for that drift chamber. This was done for all three drift chambers iteratively. This procedure of calibration of the drift chamber positions was repeated every 30 minutes of the measured data to correct for possible temperature deviation \cite{kremer,davidth}.

The magnet was operated at a current of 120 A, giving rise to a field of intensity 0.1--2 T in the region of the tracking device. The outer diameter of the coil was 61~cm and the inner diameter 36~cm. The coil was placed in a dewar filled with liquid helium surrounded by a vacuum shell enclosed in a second dewar. This dewar was filled with liquid nitrogen that reduced the rate of evaporation of liquid helium and enabled to attain a life time of about 100 hours for the superconducting magnet.

The exact position of the magnet coil, relative to the drift chamber volumes was first estimated from the drawings and then determined precisely with the tracking information from ground and flight data. The position of the magnet coil was fixed by an iterative procedure, fitting the drift chamber data using different position for the magnet. The minimization of a $\chi$\sq gave the magnet position. The magnet position was determined with an uncertainty of less than 0.1~cm \cite{davidth}.

Using the position information together with the map of the magnetic field, the rigidity of the particle was determined. 
\begin{figure}[!hbtp]
\includegraphics[width=1.\textwidth]{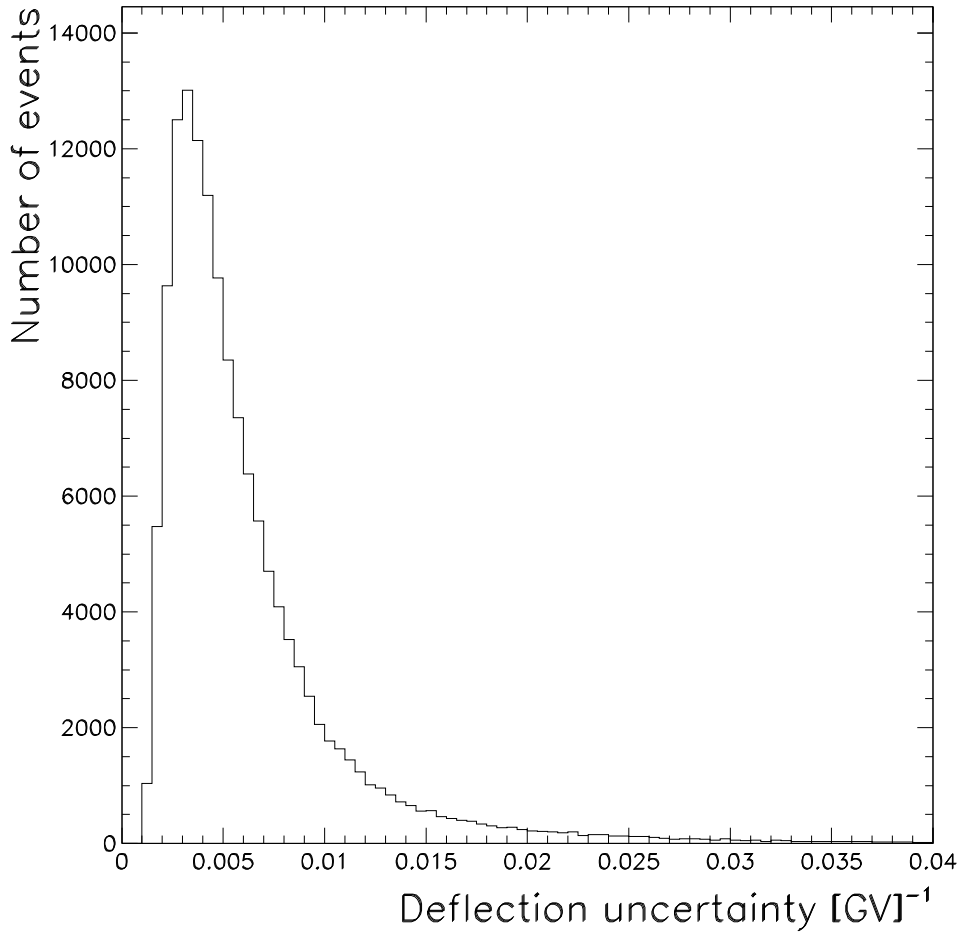}
\caption{Distribution of the deflection uncertainty for protons.}
\label{sigmadef}
\end{figure}
From the distribution of the deflection uncertainty, obtained on an event--by--event basis during the fitting procedure (see \cite{golden3}), which is shown in figure \ref{sigmadef}, an average Maximum Detectable Rigidity (MDR) of 300~GV was obtained.

%&&&&&&&&&&&&&&&&&&&&&&&&&&&&&&&&&&&&&&&&&&&&&&&&&&&&&&&&&&&&&&&&&&&&&&&&&&&&&&
\subsection{The Calorimeter}
%&&&&&&&&&&&&&&&&&&&&&&&&&&&&&&&&&&&&&&&&&&&&&&&&&&&&&&&&&&&&&&&&&&&&&&&&&&&&&&
The CAPRICE98 calorimeter configuration was the same as that in the previous CAPRICE experiments. It was designed \cite{bocciolini,ricci} to distinguish non--interacting minimum ionizing particles, hadronic and electromagnetic showers.

The calorimeter consisted of eight 48$\times$48~cm$^{2}$ silicon planes interleaved with seven tungsten converters, each one radiation length (X$_{0}$) thick. A single plane consisted of an array of 8$\times$8 silicon detectors. Each detector had a total area of 60$\times$60~mm$^{2}$ and was divided into 16 strips, each of width 3.6~mm. For every silicon plane the detectors were mounted on the two sides of a G10~motherboard with perpendicular strips to give $x$ and $y$ readout. The strips of each detectors were daisy--chained longitudinally to form a 48~cm long single strip.

This design provided a high longitudinal and transversal granularity for shower imaging. The total depth of the calorimeter was 7.2 radiation lengths and 0.33 interaction length for protons.

%##############################################################################
\section{Data analysis}
\label{datana}
%##############################################################################
The analysis was based on 21~hours of data for a total acquisition time of 67240~seconds under an average residual atmosphere of 5.5~g/cm\sq. The fractional live time during the flight was 0.4865$\pm$0.0002 resulting in a total live time ($T_{live}$)\linebreak of 32712$\pm$13~s.

Protons are the most common singly charged positive particles in the cosmic radiation. In measurements made with balloon--borne instruments there is also a small contribution of secondary particles produced in the residual atmosphere above the detector, as well as in the instrument itself. In the case of singly charged positive particles, these are mainly protons, muons, pions, and positrons. There is also a small component of primary positrons. Apart from protons, helium nuclei are the most abundant particles in the cosmic--rays; they are mainly present in the form of the $^{4}$He isotope.

The combination of detectors of the CAPRICE98 apparatus provided the redundant measurements needed for cross--checks of the in--flight detector performances. It allowed clean samples of particles to be selected by subsets of these detectors, making it possible to accurately determine the rigidity--dependent efficiency and rejection power of each individual detector and to estimate their systematic errors.

All selected singly charged particles, i.e. hydrogen nuclei, which include deuterons, were treated as being protons in this analysis. Moreover, no attempt was made in this analysis to separate $^{3}$He from $^{4}$He nuclei, and hence, all Z~$=$~2 particles were treated as being $^{4}$He.

It can be pointed out that usually the isotope discrimination can be made only over a limited energy region depending on the experimental setup and further, in most of the publications (i.e. \cite{sanuki}) relating to elemental spectra, the above procedure had been adopted. Therefore, we followed the same procedure for the CAPRICE98 experiment and this provides a meaningful comparison with other published data. Isotope abundances are being analyzed separately (i.e. \cite{vannuccini,vannuccini2}) and will be published at a later date.

%&&&&&&&&&&&&&&&&&&&&&&&&&&&&&&&&&&&&&&&&&&&&&&&&&&&&&&&&&&&&&&&&&&&&&&&&&&&&&&
\subsection{Particle Selection}
%&&&&&&&&&&&&&&&&&&&&&&&&&&&&&&&&&&&&&&&&&&&&&&&&&&&&&&&&&&&&&&&&&&&&&&&&&&&&&&
Because of the over abundance of protons and helium nuclei among primary cosmic rays the contamination of other particles in the samples was not a major issue. Hence, the selection was optimized to have an efficiency as high as possible. 

\begin{figure}[!hbtp]
\includegraphics[width=0.47\textwidth]{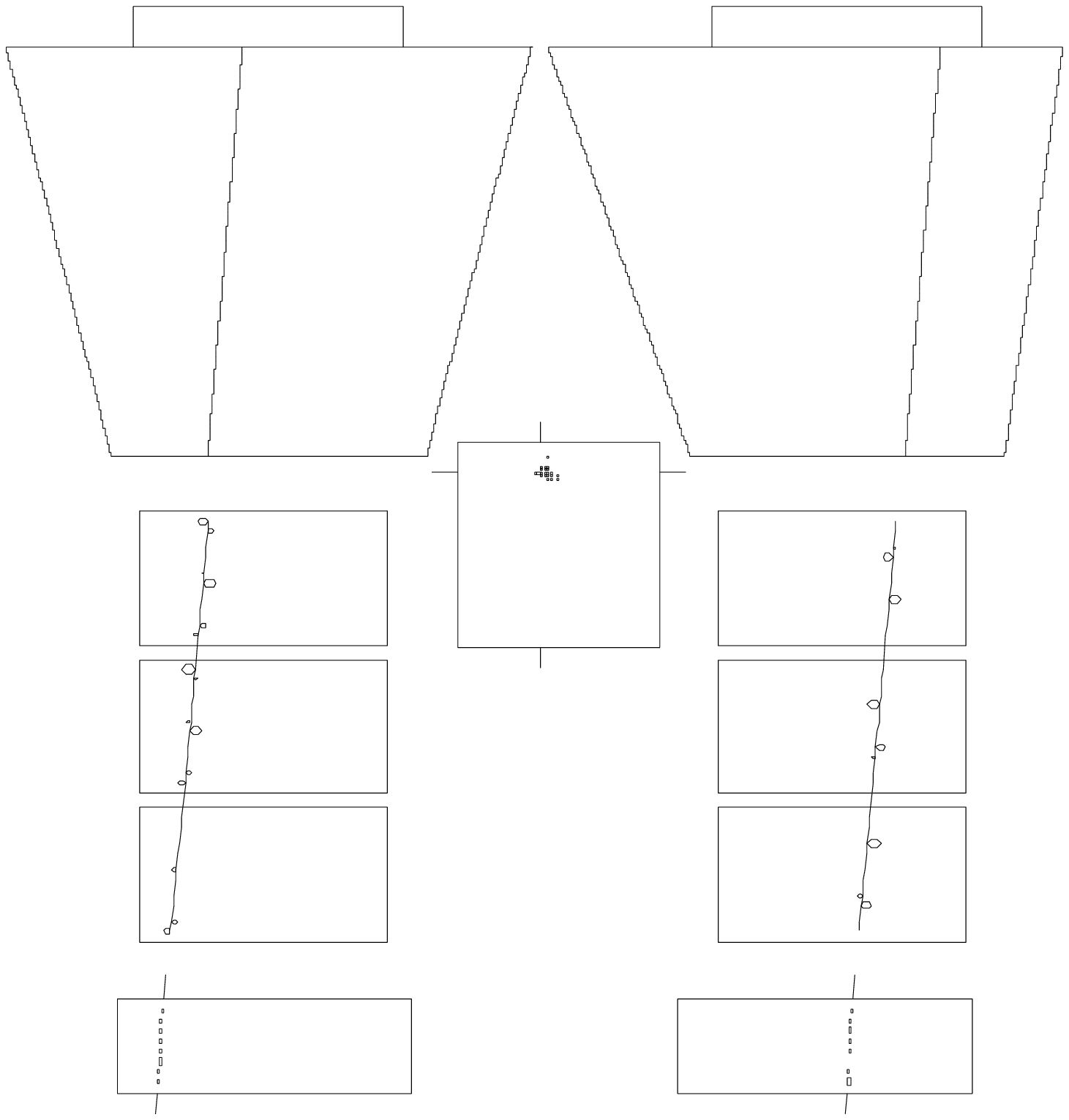}
\hspace{0.5cm}
\includegraphics[width=0.47\textwidth]{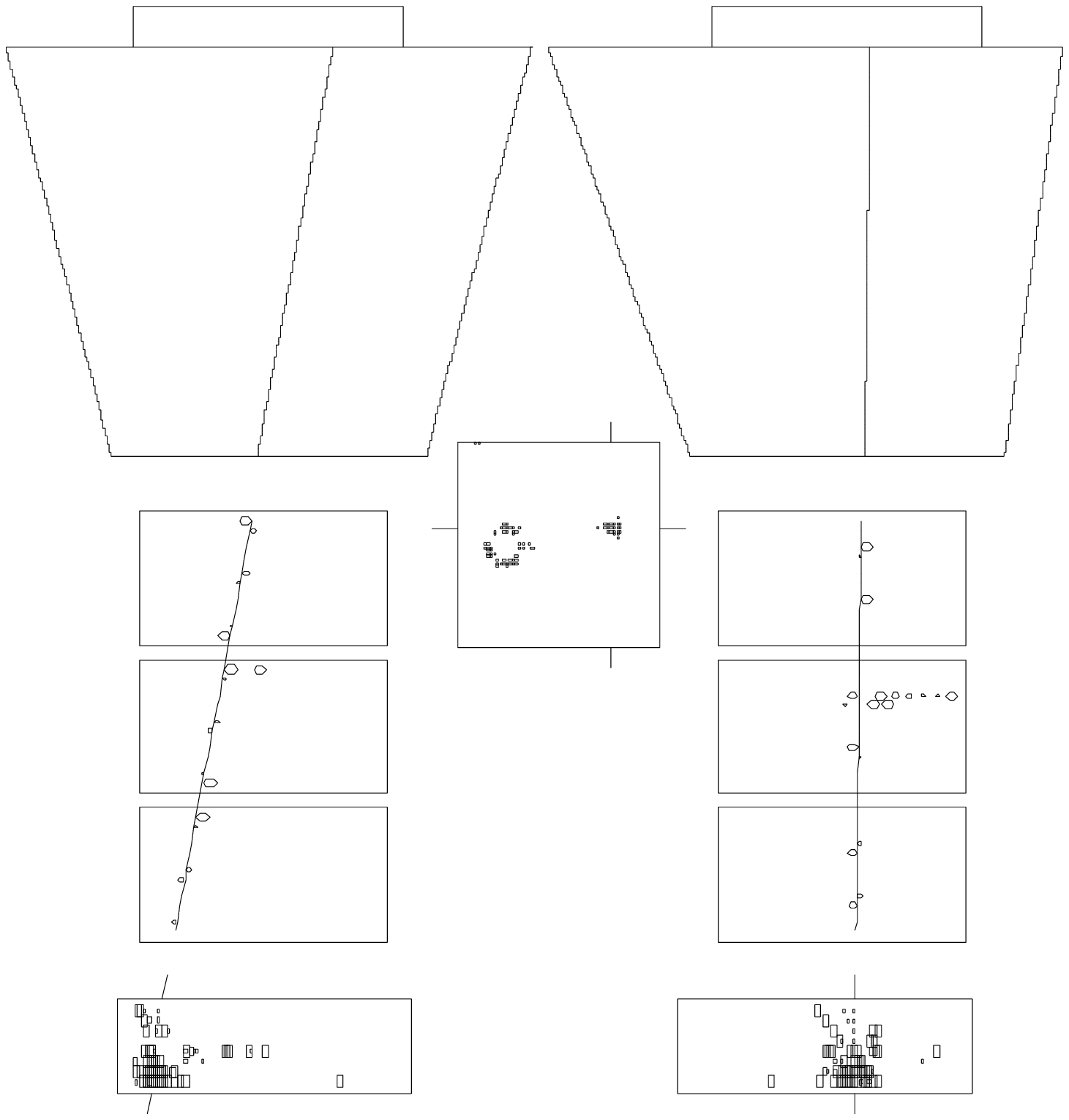}
\vspace*{0.5cm}

\includegraphics[width=0.47\textwidth]{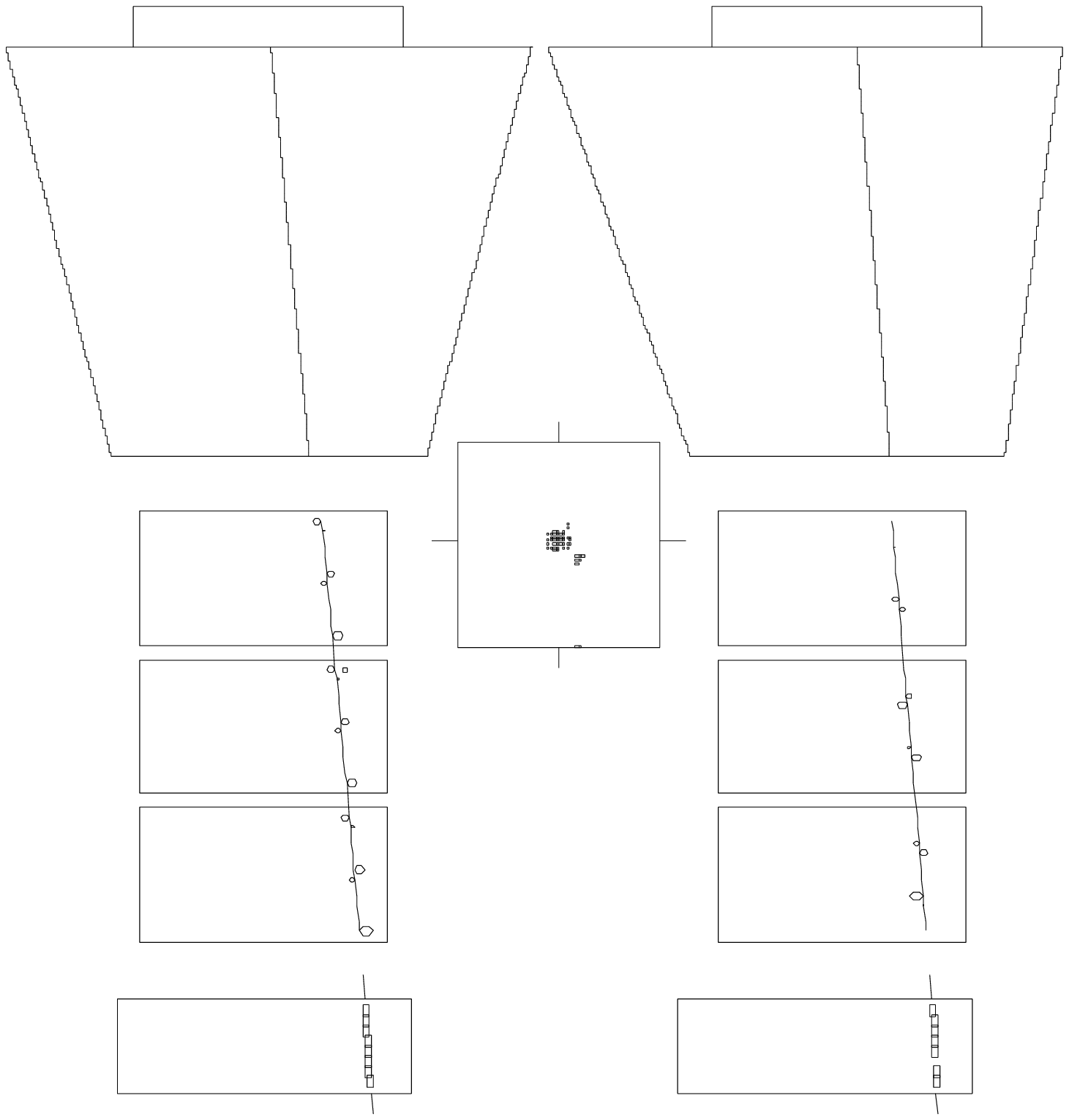}
\hspace{0.5cm}
\includegraphics[width=0.47\textwidth]{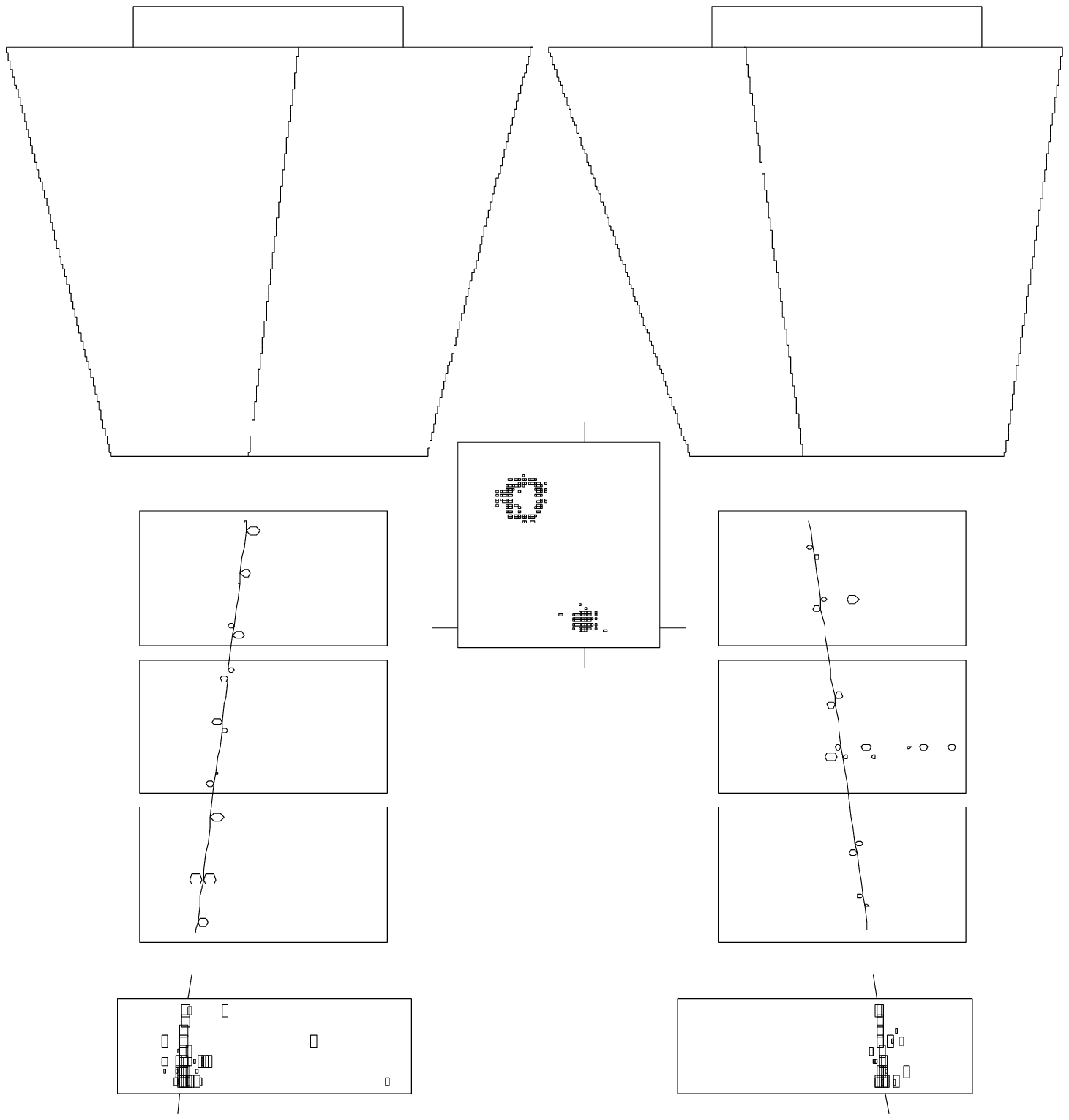}
\caption{Display of two protons and two helium nuclei traversing the CAPRICE98 apparatus. Clockwise from the upper left: a 5.6~GV non interacting proton, in the RICH only a ionization cluster is visible; a 37.8~GV interacting proton, the ionization cluster of pads can be seen well separated from the Cherenkov ring; a 82.5~GV interacting helium, notice that the ionization and Cherenkov energy released in the RICH is higher than for a singly charged particle; a 26.7~GV non interacting helium, note the signature of a helium nucleus in the calorimeter, where the size of the square is proportional to the detected energy: the ionization loss energy released is four time higher with respect to the case of a proton.}
\label{eventi}
\end{figure}

Figure \ref{eventi} shows a schematic view of two protons (top) and  two helium nuclei (bottom) in the CAPRICE98 apparatus. Each of the four sub--figures shows two panels, corresponding respectively to the $x$ and $y$ view. The RICH detector is shown at the top. A rotated view of the signals in the pad plane of the multi--wire proportional chamber is shown in the square frame in the center of the figure.  The three central boxes are the drift chambers of the tracking system. The box at the bottom shows the calorimeter information and the line drawn through all detectors represents the fitted track of the particle (note that the calorimeter is not drawn to scale). 
In the left are shown the non--interacting particles and to the right are a proton and a helium nucleus interacting in the calorimeter producing hadronic showers. 

%//////////////////////////////////////////////////////////////////////////////
\subsubsection{Tracking}
%//////////////////////////////////////////////////////////////////////////////
The tracking information was used to determine the rigidity of the particles. In order to eliminate events with more than one track in the spectrometer and to achieve a reliable estimation of the rigidity, a set of conditions was imposed on the fitted tracks:
\begin{itemize}
  \item at least 11 (out of 18) position measurements in the $x$ direction and 7 (out of 12) in the $y$ direction were required;
  \item an acceptable $\chi$\sq for the fitted track in both directions was required as well;
  \item the estimated error on the deflection had to be less than 0.02~GV$^{-1}$.
\end{itemize}

The choice of these cuts was partly based on the experience gained previously using the same tracking system \cite{boezio99b,hof96,mitchell} and it allowed a reliable estimation of the rigidity together with a high tracking efficiency, section \ref{trackeff} . The same set of conditions was used both for protons and helium nuclei.

%//////////////////////////////////////////////////////////////////////////////
\subsubsection{Time--of--flight and scintillators pulse height}
\label{tofsel}
%//////////////////////////////////////////////////////////////////////////////
The TOF information was used for selecting downward moving particles. The resolution of 230~ps, which was small compared to the flight time of more than 4~ns, assured that no contamination from albedo particles remained in the selected sample.

\begin{description}
  \item [Protons:] the pulse height information, shown in figure \ref{mip}, from the top scintillator was used to select singly charged particles as well as to reject multi--particle events coming from interactions above the top scintillator. This was done by requiring the following two conditions:

\begin{figure}[!hbtp]
\includegraphics[width=1.\textwidth]{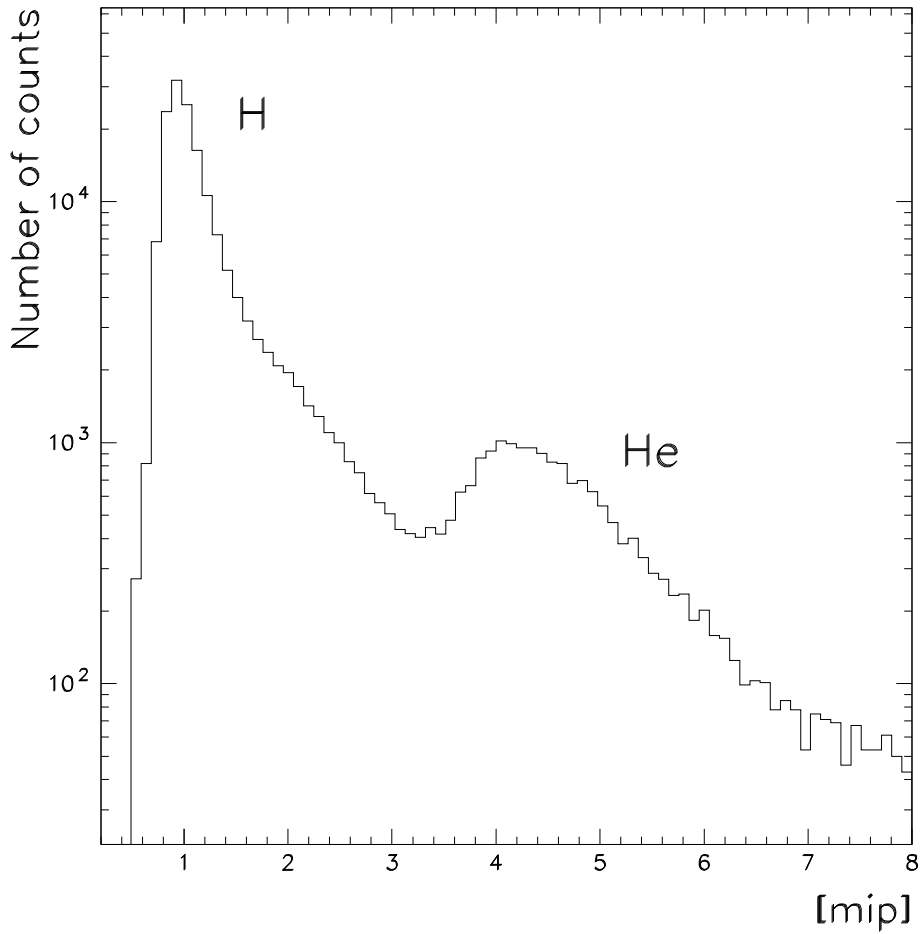}
\caption{The pulse--height spectrum, top scintillator, for a sample of events; both the singly charged particles and the helium nuclei peaks can be easily identified.}
\label{mip}
\end{figure}

  \begin{itemize}
    \item dE/dx losses in the top scintillator to be less than 1.8~mip (where a mip is the most probable energy loss for a minimum ionizing particle). This condition was chosen in order to reject about 90\% of the multi--particle events.
    \item only one paddle must be hit in the top scintillator plane.
  \end{itemize}

Protons interacting in the calorimeter could produce backscattered particles that traverse the bottom scintillator paddles giving an additional signal. None of these cases significantly affected the performance of the tracking system. Therefore, no restrictions were put on on the pulse height from the bottom scintillator. Multiple tracks were also rejected by requiring that no more than one of the two paddles in the bottom scintillator plane was hit.

  \item [Helium nuclei:] to select particle with charge two, both the top and bottom scintillators were used:
  \begin{itemize}
    \item dE/dx losses in the top scintillator should be greater than 3, but less than 7~mip;
    \item only one paddle must be hit in the top scintillator plane;
    \item dE/dx losses in the bottom scintillator should be greater than 3~mip.
  \end{itemize}
This selection criteria on the bottom scintillator was chosen in order to eliminate singly charged particles that remained in the selection after the dE/dx cut in the top scintillator. But, no upper limit for the dE/dx cut was used on the bottom scintillator in order to increase the efficiency by not discarding helium nuclei that interacted in the calorimeter and produced backscattered particles passing trough the bottom scintillator.
\end{description}

%//////////////////////////////////////////////////////////////////////////////
\subsubsection{RICH}
%//////////////////////////////////////////////////////////////////////////////
In the case of helium the RICH was used only to select helium samples for efficiency studies.

It was used as a threshold device to select protons by requiring that:
\begin{itemize}
  \item there was a good agreement between the particle impact position as determined by the RICH and the tracking system; the difference in $x$ and $y$ had to be less than 3~$\sigma$, see figure \ref{richtrack};

\begin{figure}[!hbtp]
\includegraphics[width=1.\textwidth]{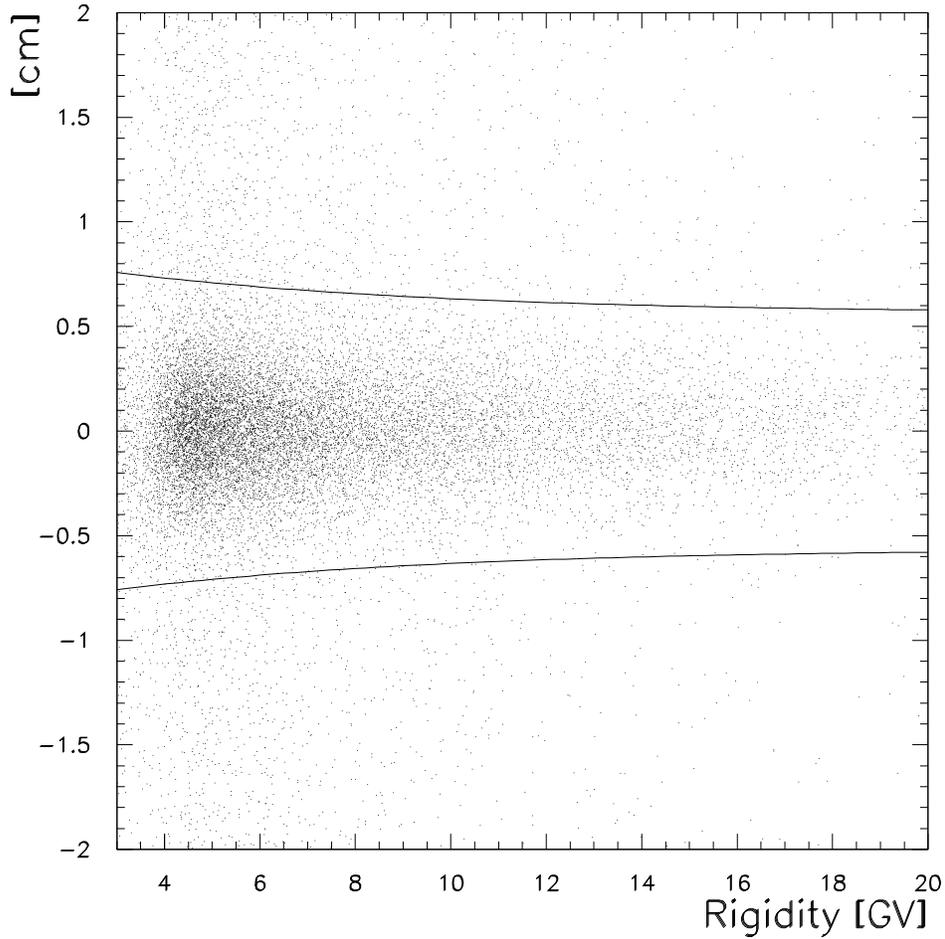}
\caption{The difference between the particle impact position on the paddle plane as determined by the RICH and the tracking system is shown as function of rigidity (in the case of the $x$ view for a sample of about 30000 events); the events between the two lines, representing the 3~$\sigma$ limits, were selected.}
\label{richtrack}
\end{figure}

  \item no pads with a saturated signal were to be found in the paddle plane outside the ionization cluster, defined as the 9~$\times$~5 pads area around the impact position measured by the tracking system \footnote{The RICH detector position resolution in the $y$ direction is better since the anode wire run along the $y$-axis in the MWPC.};
  \item no signal due to Cherenkov light was to be present for rigidities less than 14~GV.
\end{itemize}

The first two selection criteria rejected multiple tracks and events with a bad track reconstruction.

For rigidities less than 18~GV, protons were below the Cherenkov threshold and they were separated from lighter positive particles by requiring no signal due to Cherenkov light in the RICH. However, due to the finite resolution of the tracking system and the change of pressure in the RICH during the flight, the selection requirement of no Cherenkov light was used only below 14~GV, in order to maximize the efficiency. 

%//////////////////////////////////////////////////////////////////////////////
\subsubsection{Calorimeter}
%//////////////////////////////////////////////////////////////////////////////
The calorimeter provided topological information to select non--interacting particles or to discriminate between hadronic and electromagnetic showers. The measurement of the energy loss of a particle obtained with the silicon strips of the calorimeter allowed also to measure the absolute value of the charge of \linebreak non--interacting particles. However, because of the low efficiency of this selection, the calorimeter was used only to select proton and helium samples for efficiency studies of the other detectors.

%&&&&&&&&&&&&&&&&&&&&&&&&&&&&&&&&&&&&&&&&&&&&&&&&&&&&&&&&&&&&&&&&&&&&&&&&&&&&&&
\subsection{Selection efficiencies}
\label{efficiencies}
\begin{figure}[!hbtp]
\begin{center}
\includegraphics[width=0.47\textwidth]{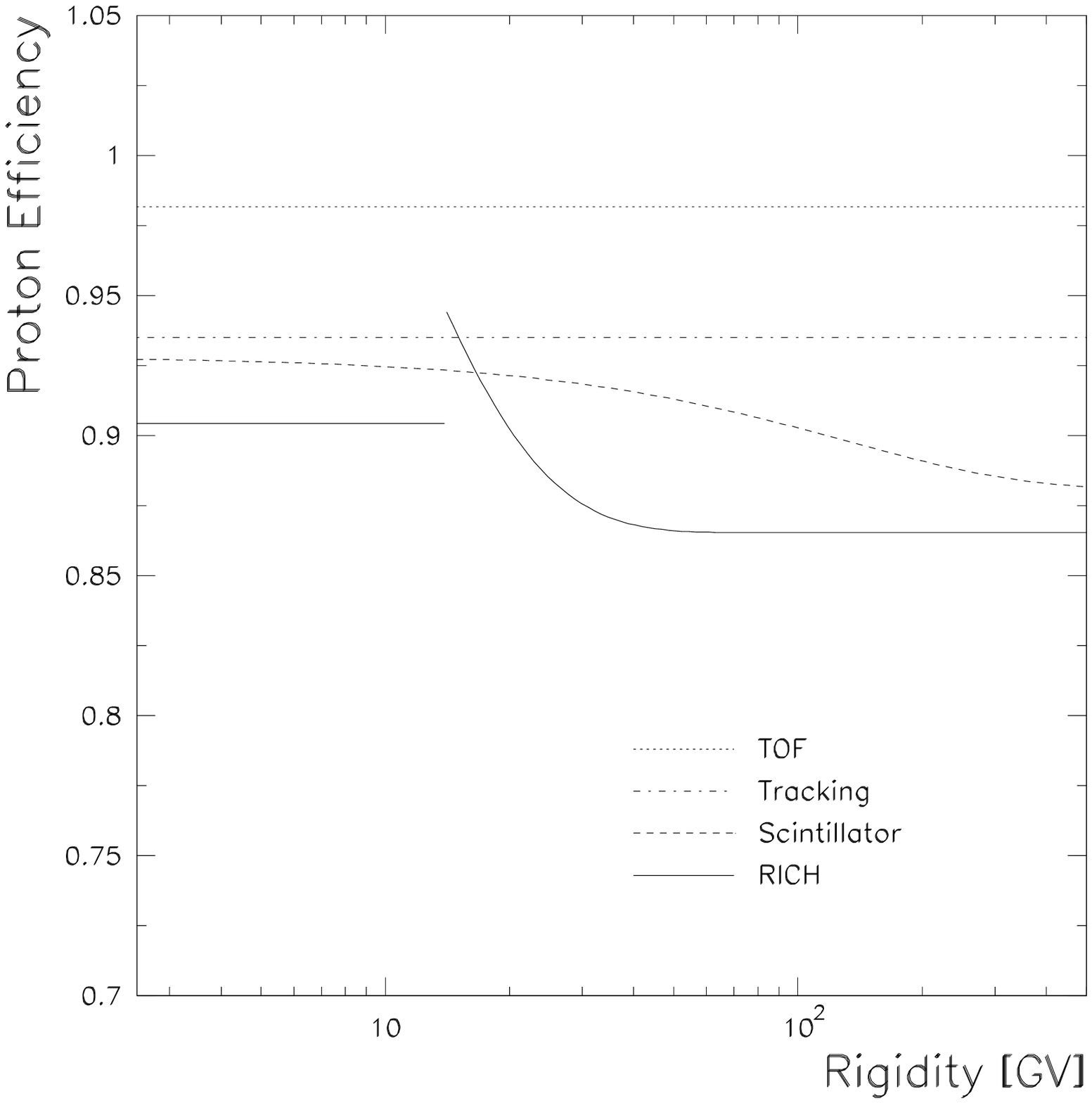}
\includegraphics[width=0.47\textwidth]{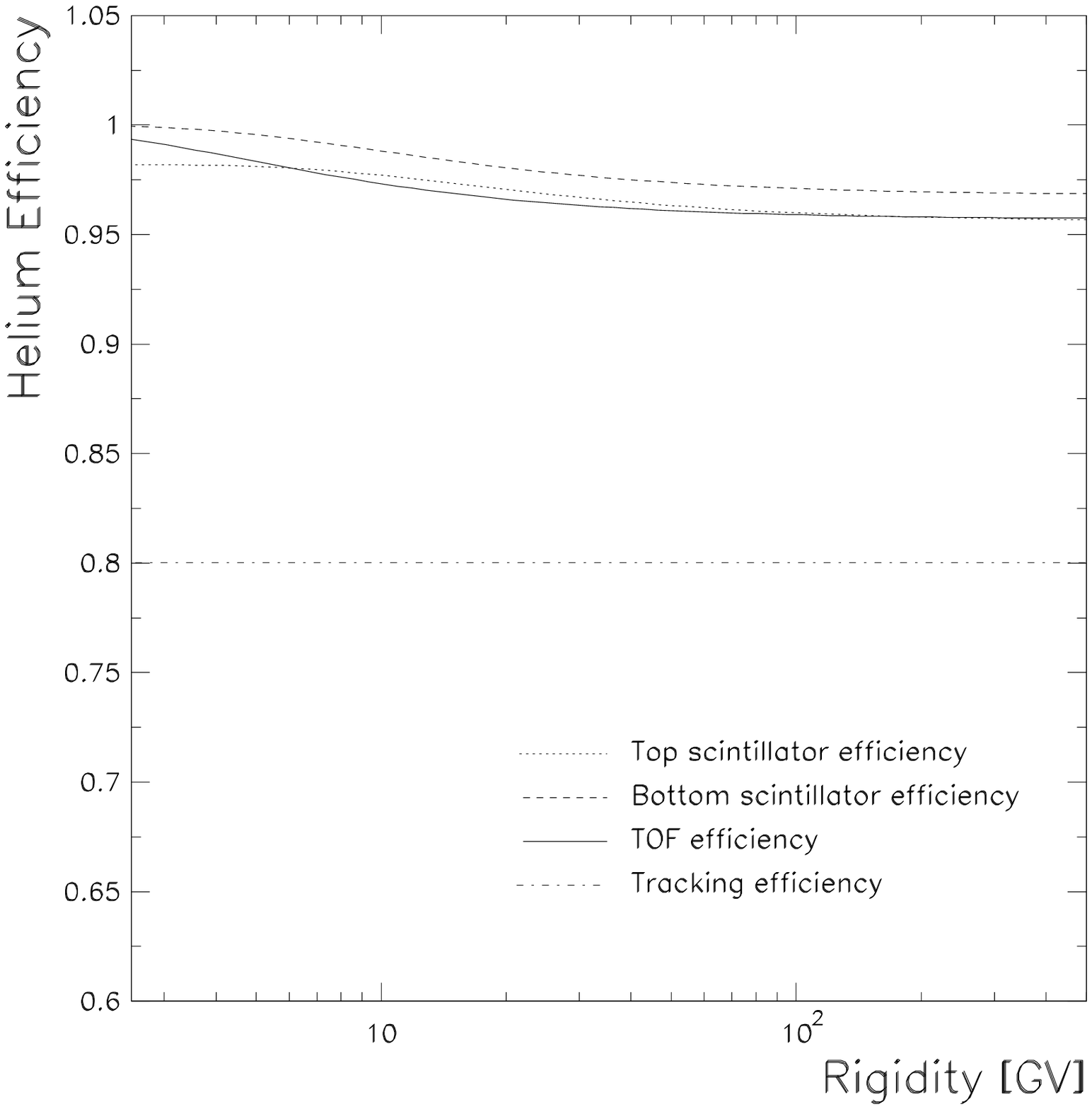}
\includegraphics[width=0.47\textwidth]{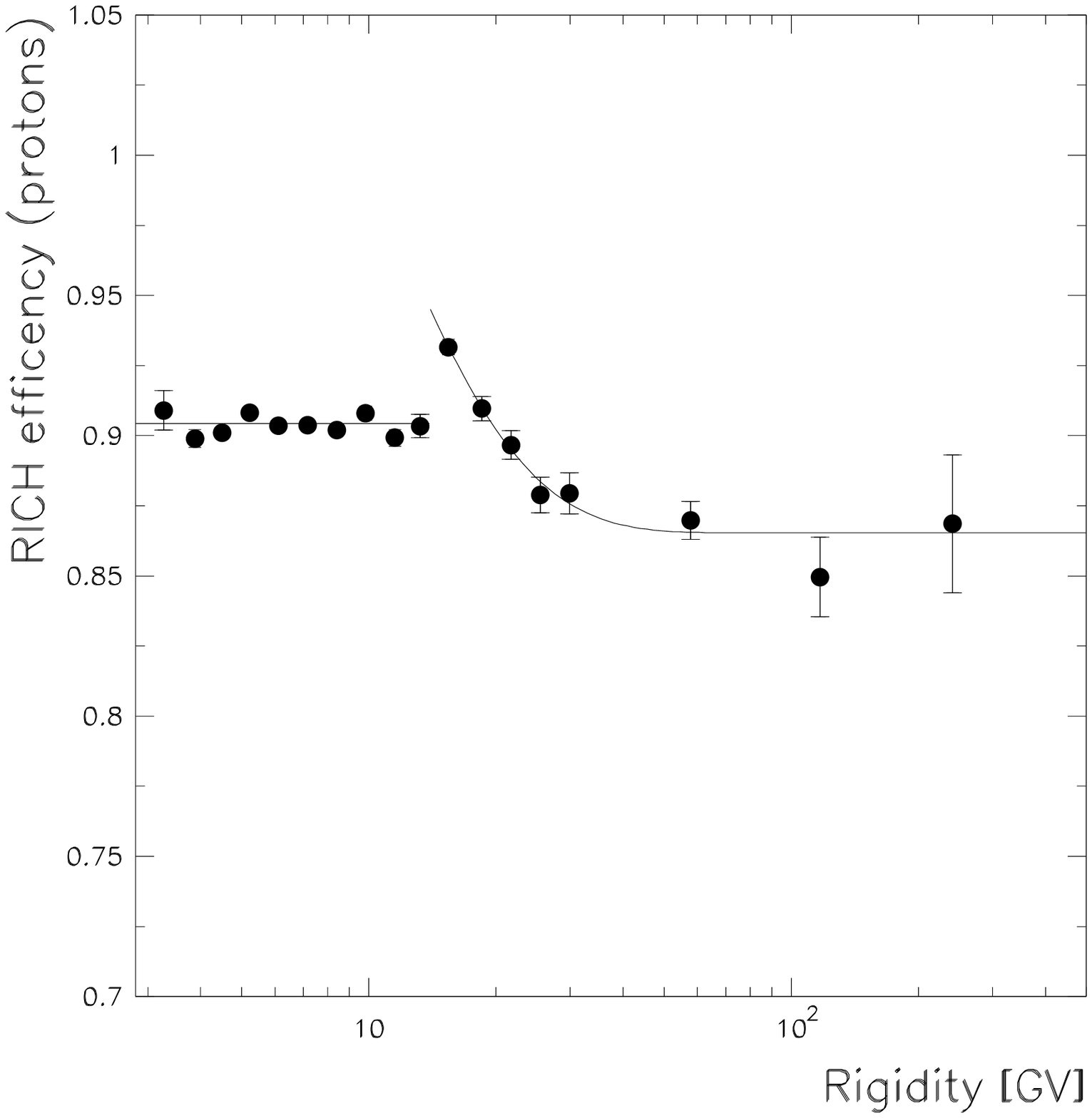}
\includegraphics[width=0.47\textwidth]{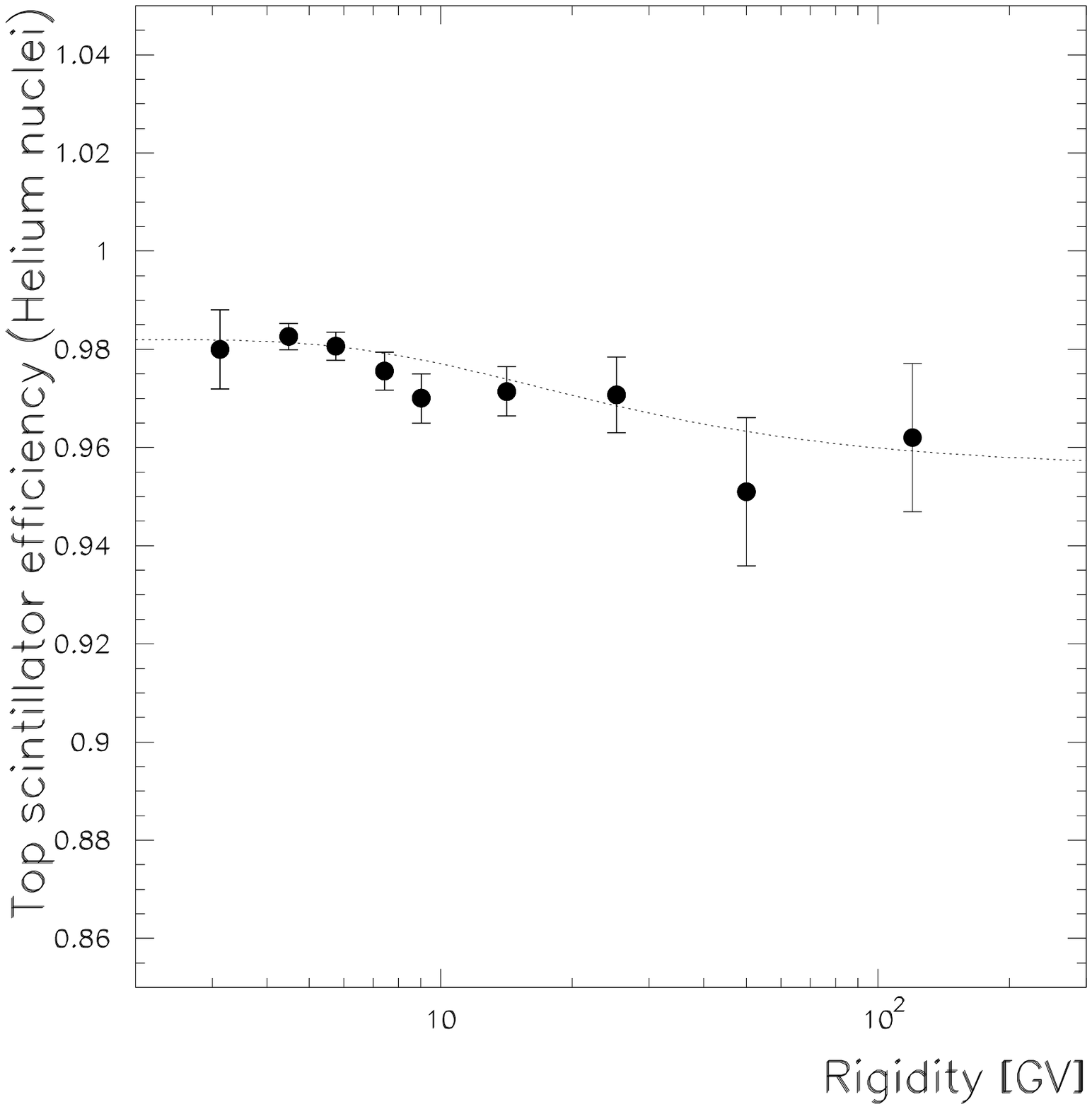}
\caption{First row: selection efficiencies as a function of rigidity for protons (left) and helium nuclei (right). The total efficiency, not shown in figures, is the product of the different functions. Second row, two examples of efficiency determination: the RICH selection efficiency for proton (left) and the top scintillator efficiency for helium nuclei (right).}
\label{ef}
\end{center}
\end{figure}

%&&&&&&&&&&&&&&&&&&&&&&&&&&&&&&&&&&&&&&&&&&&&&&&&&&&&&&&&&&&&&&&&&&&&&&&&&&&&&&
The sophisticated particle--identifying detectors used in this experiment made it possible to select clean and independent samples of different particle types to determine the efficiency of each detector. The resulting efficiencies are shown as a function of rigidity in figure \ref{ef} and are discussed in more detail below.

The efficiency of each detector was determined as a function of rigidity in a number of discrete bins. The efficiency was then parametrized to allow an interpolation between bins. This parametrization could introduce a systematic error on the efficiency of each detector. Since the parameters were correlated, the error on the efficiency was obtained using the error matrix of the fit for each detector when correcting the measured flux for the detector efficiencies. Then, assuming that the resulting systematic errors associated with different detector efficiencies were uncorrelated, they were quadratically summed.

%//////////////////////////////////////////////////////////////////////////////
\subsubsection{Tracking efficiency}
\label{trackeff}
%//////////////////////////////////////////////////////////////////////////////
The tracking efficiency was obtained from test samples of protons and helium nuclei selected by combining the information from the other detectors. 

\begin{description}
  \item[Protons:] two different and independent methods were used to determine the tracking efficiency and both of them were independent of the drift chamber tracking system. They were first tested with muons from ground data since it was expected, from previous experience \cite{weber}, that muons and protons had the same tracking efficiency above a few GeV. Then the final efficiency was obtained with float data only.

  \begin{itemize}
    \item The first method, called ``RICH method'', used the RICH detector to get the rigidity from the Cherenkov angle measurement with the help of an extrapolated straight track from the calorimeter \cite{davidth}. 

At ground the rigidity was derived from the reconstructed Cherenkov angle measured by the RICH detector assuming the mass of a muon. The composition of cosmic rays at ground is dominated by muons, but there is also a non--negligible fraction of electrons below 1~GeV \cite{boezio98} and of protons below about 15~GeV \cite{kremer,golden}. The calorimeter selection was able to reduce the electron contamination to a negligible amount and protons were not a source of contamination since they did not produce Cherenkov light below 18~GV.

At float, protons are the most abundant component, but there are also singly charged particles, such as muons, electrons and positrons. This led to an ambiguity in the definition of the tracking efficiency sample. Since the rigidity determination was based on the velocity calculated from the Cherenkov angle by assuming the mass of the proton, this method gave the same rigidity for particles with a different mass having the same measured velocity. Electrons and positrons were easily rejected by the calorimeter, but non--interacting protons were undistinguishable from muons. Hence the efficiency sample could contain muons and a small amount of pions, with rigidity greater than 2~GV along with protons of rigidity greater than 18~GV. As a result, if high energy muons and protons have different tracking efficiency, this could bias the result.

    \item The second method is called ``no--DC method'' \cite{kremer}, because it does not use the drift chambers for track reconstruction, but uses the same tracking routines for the complete CAPRICE98 tracking system. Instead of the wire planes of the drift chambers, it uses the particle positions as determined by (a) the pad plane of the MWPC of the RICH detector, (b) the two planes of the scintillation detectors of the TOF system, which make use of the time difference between the signals recorded by the photomultipliers on either end of the scintillators (only in the $x$ direction), and (c) at least five layers in each view from the calorimeter (with a maximum of 8 layers per direction). Hence, this method used from a minimum of 8 (6) to a maximum of 11 (9) points for the track reconstruction in the $x$ ($y$) direction. From the track routine, the particle track was reconstructed using the above set of particle positions and the estimated MDR was found to be about 5~GV.

At ground negative muons were selected requiring negative deflection from the fit, no shower in the calorimeter, and an ionization typical of a minimum ionizing particle in the TOF scintillators. Consistency between the velocity measured by the TOF, and the expected Cherenkov signal in the RICH detector for a muon was also required.

At float the proton tracking efficiency could be directly measured. Charge one particles were selected using the dE/dx signal in the TOF scintillators, the velocity derived from the TOF measurements and requiring no Cherenkov light in the RICH detector. This selection could be contaminated by muons between 1.5 and 2~GV and by pions between 1.5 and 3~GV. 
  \end{itemize}
\vspace{0.5cm}
The two methods at ground gave the same results between 0.2 and 10~GV.
\linebreak

With flight data, the first method sampled the tracking efficiency of protons above 18~GV (because of the gas--RICH threshold) while the second below 10~GV. From previous experience with a similar tracking system \cite{boezio99b} the proton tracking efficiency was expected to reach a plateau above 2~GV but in this case the two methods differed by $\sim$2.5\%. However, the second method could be biased by the contamination of secondary low energy protons that would not affect the first method \cite{davidth}. Since the geomagnetic cut--off was at about 4.3~GV, the bulk of the singly charged particles was above 4.5~GV, primary protons, and below 1~GV, mostly secondary protons and muons. The region between 2 and 4~GV was, in proportion, highly depopulated. Since the MDR of the no--DC method was only 5~GV, even a small fraction of the low energy protons which spilled over to the rigidity bin above 2~GV could bias the sample, giving a smaller efficiency. This would affect only the proton efficiency of the second method because the no--light condition in the RICH only implied that protons had rigidities lower than 18~GV. The same does not apply at ground to the second method since the matching condition between the expected and the measured Cherenkov signal in the RICH detector was required. In the case of protons, the contamination of the no--DC sample was examined using a sample with the rigidity measured by the drift chamber tracking system. This showed that a contamination of about 10\% was present in the sample below 2~GV \cite{davidth}. Another clue to a contamination in the no--DC sample of protons is given by the helium nuclei analysis: in that case, the results from the two methods were in a very good agreement (about 1\% difference) because there was no contamination in both the samples, as the dE/dx selection rejected all singly charged particles.

Thus, the first method was used to obtain the efficiency of the tracking selection in all the rigidity range, from 4 to 350~GV, and it was constant at 93.5$\pm$0.5\%. This is shown by the dotted--dashed line in the upper part of figure \ref{ef}. However, the difference found between the proton tracking efficiency with the two methods was considered as a possible systematic uncertainty and 2.5\% systematic uncertainty was included in the flux calculation for rigidities between 4 and 20~GV.
    
  \item[Helium nuclei:] the same two methods were used also in the case of helium nuclei in the rigidity range 40~GV to 200~GV (first method) and from 0.5 to 15~GV (second method). In this case the two methods agreed and the tracking efficiency for helium nuclei was found to be constant at 80$\pm$1\%; the dotted--dashed line in the bottom of figure \ref{ef} shows this result. The difference between helium nuclei and protons is due to the higher ionization of the doubly charged particles causing cross--talk effects, and resulting in the loss of efficiency in the drift chamber.
\end{description}

%//////////////////////////////////////////////////////////////////////////////
\subsubsection{Scintillator efficiency}
%//////////////////////////////////////////////////////////////////////////////
\begin{description}
  \item [Protons:] a sample of protons selected with the RICH and the calorimeter was used to determine the TOF selection efficiency. The RICH was used to select protons above the threshold by requiring that the reconstructed Cherenkov angle should not deviate by more than 3~$\sigma$ from the expected Cherenkov angle for protons. This criterion, combined with the containment condition and the matching condition between the ionization cluster position and the impact position determined by the tracking system, allowed to reject all the up--going particles. dE/dx measurements from the first two planes of the calorimeter were used to reject helium nuclei. The TOF selection efficiency for protons is shown by the dotted line in figure \ref{ef}.  

The top scintillator efficiency for selecting singly charged particle, is shown by the dashed (``Scintillator'') curve in figure \ref{ef}. This was determined by using a sample of protons selected with the bottom scintillator and the TOF. A cross--check was made by selecting singly charged particles using the dE/dx measurements in the calorimeter.

  \item [Helium nuclei:] the TOF efficiency is shown by the solid curve in figure \ref{ef}. This was determined by selecting a sample of the helium nuclei using the dE/dx signals from the top two silicon detectors in the calorimeter.

The same particle selection using the calorimeter and the TOF selection were then used to determine the top scintillator efficiency; this is shown by the dotted curve in figure \ref{ef}.

To determine the bottom scintillator efficiency, a sample of helium nuclei was selected with the first two planes of the calorimeter, the top scintillator and the ionization energy released in the pad plane of the RICH. The estimated scintillator efficiency is shown by the dashed curve in figure \ref{ef}.
\end{description}

As can be noticed in figure \ref{ef}, the dE/dx efficiency is higher for helium nuclei than for protons. This is due to different selections applied to helium nuclei and  protons. Interactions in the payload produce singly charged multi--particles events that were rejected for protons by requiring the energy loss in the top scintillator to be less than 1.8 mip, a condition stricter with respect to the helium nuclei. The dependence on the rigidity is due to the relativistic rise of the energy loss function. 

%//////////////////////////////////////////////////////////////////////////////
\subsubsection{RICH efficiency}
%//////////////////////////////////////////////////////////////////////////////
The efficiency of the RICH was determined only for protons, since the RICH is not used in the selection of helium nuclei.

The proton sample was selected by the TOF, the dE/dx in the scintillators, and the calorimeter; the calorimeter was used to select interacting protons, in order to reject muons from the sample. The resulting efficiency is shown in figure \ref{ef} by the solid curve.

\begin{figure}[!hbtp]
\includegraphics[width=1.\textwidth]{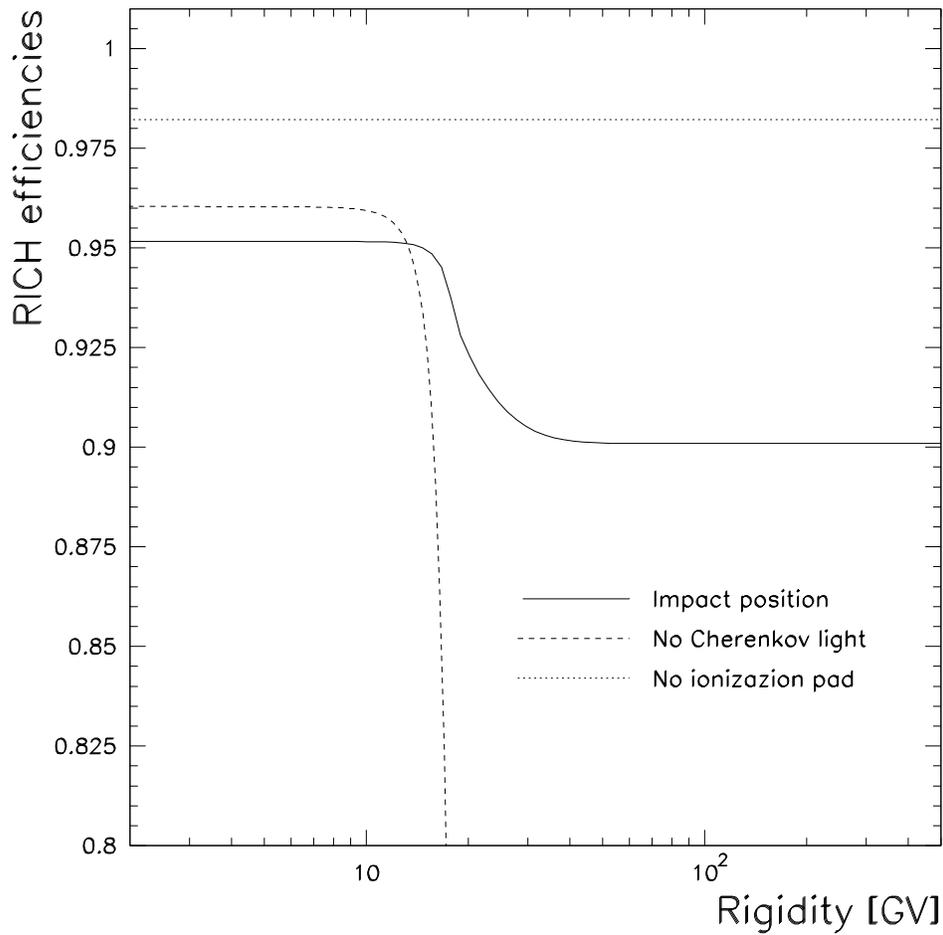}
\caption{Proton selection efficiencies as a function of rigidity for the RICH. Since the ``no--ionization'' and the ``impact position'' conditions are correlated the total efficiency for the RICH is higher than the product of the three partial efficiencies.}
\label{richef}
\end{figure}

Figure \ref{richef} shows the RICH efficiency decomposed in its three components:
\begin{itemize}
\item the efficiency obtained by requiring agreement between the particle impact position measured by the tracking system and the RICH (solid curve); 
\item the efficiency for no signal due to Cherenkov light (dashed curve);
\item the efficiency for requiring no ionization signal found outside the ionization cluster (dotted line).
\end{itemize}
Notice that the ``no--ionization'' and the ``impact position'' conditions are correlated since the ionization cluster is determined using the impact position measured by the tracking system. Because of this reason, the total efficiency for the RICH shown in figure \ref{ef} is higher than the product of the three efficiencies shown in figure \ref{richef}. 

It can be noticed that the ``impact position'' efficiency decreases when protons start to emit Cherenkov light. In fact, if the ionization cluster falls on the border of the pad plane, part of the ionization energy is lost and the Cherenkov light distributed over a cluster of paddles can be erroneously confused with the ionization energy. In this case the matching conditions between the ionization cluster and the impact position determined by the tracking system will not be satisfied.

The ``no--ionization'' condition does not depend on rigidity and the emission of Cherenkov light does not affect this selection since only the ionization by a charge particle could give a saturated signal.

The efficiency for ``no light'' condition decreases when protons start to emit Cherenkov light. Below the Cherenkov threshold, the efficiency is still lower than one because of two different effects. The first one is due to the finite resolution of the spectrometer, by which some protons with a rigidity greater than the threshold for Cherenkov emission contaminate the sample below this threshold. The second effect is due to the erroneous reconstruction of the Cherenkov angle, made by using pads lying just outside the ionization cluster or noisy pads.

%&&&&&&&&&&&&&&&&&&&&&&&&&&&&&&&&&&&&&&&&&&&&&&&&&&&&&&&&&&&&&&&&&&&&&&&&&&&&&&
\subsection{Geometrical factor}
%&&&&&&&&&&&&&&&&&&&&&&&&&&&&&&&&&&&&&&&&&&&&&&&&&&&&&&&&&&&&&&&&&&&&&&&&&&&&&&
The spectrometer accepted particles with a zenith angle less than 14 degrees. The average angle was 8 degrees.

The geometrical factor was obtained with Monte~Carlo techniques \cite{sullivan}; the simulation implemented the same track--fitting algorithm used in the analysis to trace particles through the spectrometer. The geometrical factor ($G$) was found to be constant 155.0$\pm$1.1~cm\sq~sr in the rigidity range 4--350~GV both for protons and helium nuclei since the acceptance conditions were the same in the two cases.

%&&&&&&&&&&&&&&&&&&&&&&&&&&&&&&&&&&&&&&&&&&&&&&&&&&&&&&&&&&&&&&&&&&&&&&&&&&&&&&
\subsection{Payload and atmospheric corrections}
%&&&&&&&&&&&&&&&&&&&&&&&&&&&&&&&&&&&&&&&&&&&&&&&&&&&&&&&&&&&&&&&&&&&&&&&&&&&&&&
The number of detected particles was corrected in order to compensate for the fraction of events lost due to interactions in the payload. 

To reach the tracking system of the spectrometer, the particles had to first go through the aluminium shell of the payload and the RICH detector, and then through the top scintillator of the TOF system. It was assumed that all particles that interacted above the tracking system were rejected by the selection criteria. The probability of an interaction in the material of the drift chamber, that would not be rejected by the tracking system conditions, was negligible. The data were corrected for these losses with multiplicative factors, using the expression for the interaction mean free path for the different materials in the detectors given by Stephens \cite{stephens}. The correction is about 10\% (see tables \ref{tabcorr} and \ref{tabcorrhe}).

% TABLE 1 / TABLE 2

Further corrections were made in order to propagate the flux values estimated at the top of the payload back to the top of the atmosphere. Particles produced in the atmosphere above the detector must be subtracted, the fraction of events lost because of interactions must be compensated and the energy lost by ionization during propagation has to be taken into account. In the case of protons we used the calculation made by Papini, Grimani, and Stephens \cite{papini} and in the case of helium we used the calculation made by Vannuccini \cite{vannuccini3}. The atmospheric correction was about 10\%.

Tables \ref{tabcorr} and \ref{tabcorrhe} report the number of events observed at the spectrometer and how it changes by applying the payload and atmospheric corrections. The first column represents the rigidity bins at the spectrometer; it corresponds to different intervals at the top of the atmosphere due to ionization losses. The second column contains the selected number of events. By dividing this number by the total efficiency and by accounting the payload correction (not shown in tables) it is possible to obtain the third column (in the case of the first row of table \ref{tabcorr}, i.e., $9106/(0.935*0.982*0.904*0.927)\simeq 11843*1.079  = 12779$). In the fourth column is listed the number of lost events due to interactions in the residual atmosphere. By summing these numbers to the third column and by subtracting the number of secondary protons produced in the atmosphere (not shown in tables) it is possible to obtain the extrapolated number of events at the top of the atmosphere, fifth column (always in the case of the first row of table \ref{tabcorr}, $12779+1570-796=13553$). 

%&&&&&&&&&&&&&&&&&&&&&&&&&&&&&&&&&&&&&&&&&&&&&&&&&&&&&&&&&&&&&&&&&&&&&&&&&&&&&&
\subsection{Geomagnetic transmission correction}
%&&&&&&&&&&&&&&&&&&&&&&&&&&&&&&&&&&&&&&&&&&&&&&&&&&&&&&&&&&&&&&&&&&&&&&&&&&&&&&
The transmission of the particles through the Earth's magnetic field had to be taken into account in order to get the fluxes at the top of the atmosphere. The average value of the vertical cut--off rigidity was about 4.3~GV. However, this cut--off is not a sharp one below which all particles are deflected back to space and above which all particles arrive at the apparatus. In fact, around the geomagnetic cut--off the particles are partially transmitted through the Earth magnetic field. Furthermore, the penumbral bands define forbidden bands of rigidity, which vary with arrival direction, time and geographical location. In this analysis all these effects are represented by a single transmission function, which was derived from the experimental data.

We found that the CAPRICE94 \cite{boezio99b} and CAPRICE98 helium spectra above about 10~GeV/n are nearly identical in shape and the absolute fluxes differ by less than 4\%, a good agreement considering statistical errors. Moreover, the solar modulation during the two balloon flights was also very similar. The values from the neutron monitor counter CLIMAX \cite{simpson} were 415600 counts/hour and 417000 counts/\linebreak hour at the time of the CAPRICE94 and CAPRICE98 flights, respectively. The CAPRICE94 experiment took place in North Canada at an average geomagnetic cut--off of about 0.5~GV. Hence, the effects of the geomagnetic field on the CAPRICE94 proton and helium nuclei spectra were negligible above 1~GV. Consequently, the transmission function was defined as the ratio between the helium fluxes measured by CAPRICE98 and CAPRICE94.

\begin{figure}[!hbtp]
\includegraphics[width=1.\textwidth]{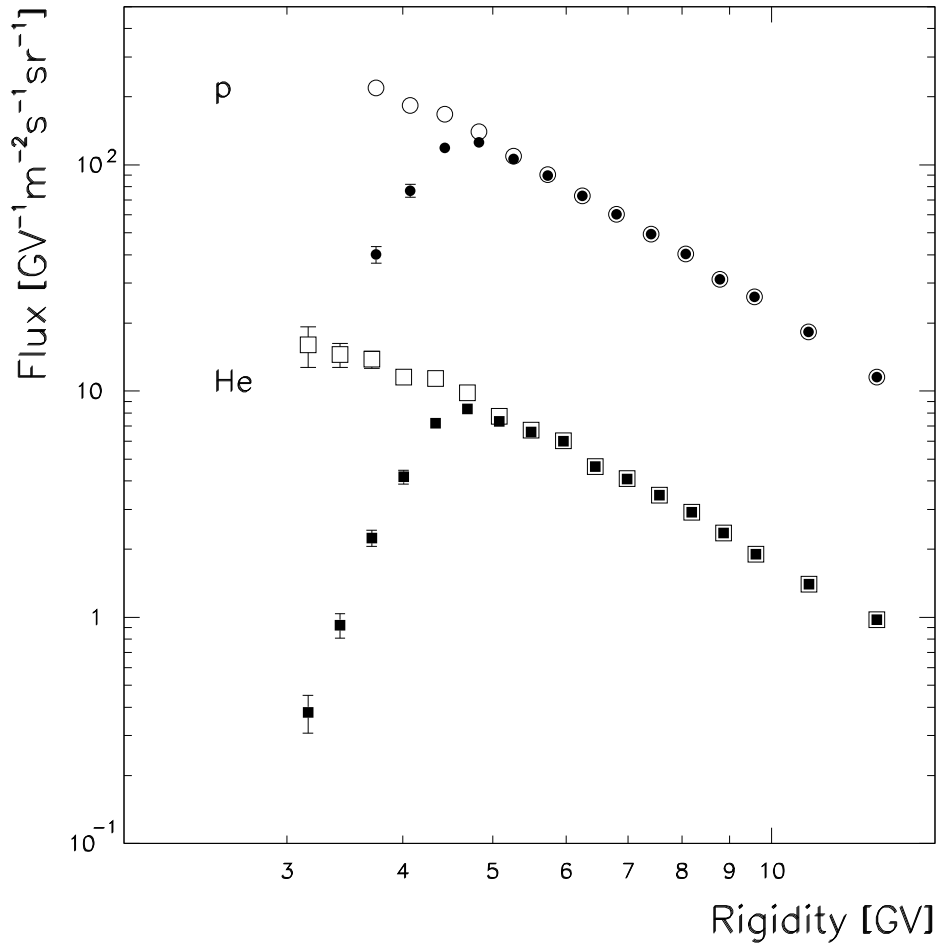}
\caption{Effects of the geomagnetic transmission correction on the proton and helium nuclei fluxes. Solid symbols represent the fluxes without correction while open symbols are fluxes after the correction.}
\label{efftrf}
\end{figure}

We have made use of this experimentally derived transmission function to obtain the flux of proton and helium nuclei. In figure \ref{efftrf} we have shown the fluxes of proton and helium nuclei as a function of rigidity before and after correcting for the transmission by solid and open symbols respectively, and one can notice that the value of the transmission function decreases from one only below 5~GV.

%&&&&&&&&&&&&&&&&&&&&&&&&&&&&&&&&&&&&&&&&&&&&&&&&&&&&&&&&&&&&&&&&&&&&&&&&&&&&&&
\subsection{Contaminations}
%&&&&&&&&&&&&&&&&&&&&&&&&&&&&&&&&&&&&&&&&&&&&&&&&&&&&&&&&&&&&&&&&&&&&&&&&&&&&&&
The contamination of muons, pions and heavy elements in the proton sample was carefully studied, as well as the singly charged particle contamination in the helium nuclei sample. As said before, no attempt was made to separate deuterons from protons and $^{3}$He from $^{4}$He.

\begin{description}
\item[Z$=$1 particle contamination in the proton sample:] the muon and pion contamination in the proton sample was studied using negatively charged muons recorded at ground before the flight. The surviving fraction of muons and pions after applying the proton selection criteria was found to be (0.36$\pm$0.07)\% in the energy range between 3 and 14~GV. Above this rigidity, the no--light selection in the RICH was not used and hence, muons and pions were included in the proton sample above 14 GV. We neglected this small contamination in our analysis. The positron contamination in the proton sample above 3~GeV was negligible and it was neglected in our analysis.

\item[Z$>$1 particle contamination in the proton sample:] heavier particles, mainly \linebreak helium nuclei, were efficiently rejected by the scintillator dE/dx selection criteria. A test sample selected by the calorimeter shows that the contamination of heavier elements in the proton sample was less than 0.2\% independent of rigidity.

\item[Z$\ne$2 particle contamination in the helium nuclei sample:] the contamination of \linebreak charge one particles in the analysis of helium nuclei was studied by selecting a sample of Z$=$1 particles with the calorimeter. The dE/dx scintillator selection applied to this sample shows a negligible ($<$ 0.1\%) contamination. The contamination of higher Z particles was also negligible, because of the condition on the pulse heigh information on the top scintillator and because of the small amount of these particles in the cosmic--rays.
\end{description}

%&&&&&&&&&&&&&&&&&&&&&&&&&&&&&&&&&&&&&&&&&&&&&&&&&&&&&&&&&&&&&&&&&&&&&&&&&&&&&&
\subsection{Systematic uncertainties}
\label{syserr}
%&&&&&&&&&&&&&&&&&&&&&&&&&&&&&&&&&&&&&&&&&&&&&&&&&&&&&&&&&&&&&&&&&&&&&&&&&&&&&&
Systematic errors originating from the determination of the detector efficiencies have already been discussed in section \ref{efficiencies} and they have been included in evaluating the errors associated with flux values, which are shown in the figures and in the tables.

Other possible sources of systematic errors are listed below and are also included in the table \ref{ester}:
% TABLE 3

\begin{description}
\item[Trigger efficiency:] the trigger efficiency was studied during the pre--flight preparations with a system providing particle coincidence between two scintillators, one placed above the top and the other below the bottom TOF scintillators. The performance of the trigger system during the flight was also studied by comparing the experimental spatial distribution of triggers with the distribution given by the same simulation, which was used for the geometrical factor calculation, and an excellent agreement was found. From the pre--flight measurements, the efficiency was found to be close to 100\% with an uncertainty of about 2\%.

\item[Geometrical factor:] the method for calculating the geometrical factor used in this work was compared with two other techniques in the CAPRICE94 analysis \cite{boezio99b}, and it was found to be in agreement within 2\% above 0.5~GV. Considering the similar geometrical configuration of CAPRICE98, it can be concluded that the uncertainty in the geometrical factor was about 2\%.

\item[Atmospheric secondaries:] the estimate of the atmospheric secondaries was made by repeating the calculations of Papini et al. \cite{papini} using the spectra obtained in this experiment. We do not expect the systematic errors in this estimate to be larger than 10\%, that gives a negligible ($<$ 1\%) influence on the results. The atmospheric secondaries were also affected by the uncertainty in the residual atmosphere above the gondola. The mean residual atmosphere was measured to be 5.5~g/cm\sq by a pressure sensor owned and calibrated by the CAPRICE collaboration; the variation in altitude during the flight introduces a deviation from the mean of less than 0.3~g/cm\sq. The pressure was also measured by a detector owned and calibrated by the National Scientific Balloon Facility (NSBF). The NSBF pressure data were about 15\% lower (i.e. about 4.6~g/cm\sq) at float than the ones measured by our sensor. This introduces an uncertainty in the fluxes of less than 1.5\% independent of energy. 

\item[Losses in the atmosphere and payload:] the numbers of particles measured at the spectrometer were corrected for losses in the spectrometer and in the atmosphere. 10\% uncertainty in the cross sections used in these calculations leads to a systematic error on the proton and helium nuclei fluxes of about 2\% from losses in the payload and the residual atmosphere. An additional uncertainty of 1\% should be considered due to the uncertainty in the atmospheric depths and the consequent effect on the losses in the atmosphere.

\item[Spectrometer resolution:] the CAPRICE98 results were not deconvolved for the effect of the intrinsic spectrometer resolutions since the effect on the measured fluxes is smaller than the statistical errors. Instead a systematic error has been included that accounts for the finite spectrometer resolution. Figure \ref{sigmadef} shows the distribution of the deflection uncertainty for protons. The same method used for the CAPRICE94 experiment, see \cite{boezio99b}, was applied and, assuming as input a power law spectrum with $\gamma=-2.7$, the systematic error found is less than 2\% up to 200~GV and reaches 5\% at 350~GV on the fluxes.
\end{description}

Other systematic uncertainties due to the tracking system can arise during the flight. For example, the variation of the temperature can cause a misalignment of the drift chambers, resulting in an offset in the deflection measurement. Another possible source of systematic errors could be a wrongly mapped magnetic field or an error in the centering of the magnet with respect to the drift chambers. These effects would result again in a wrong measurement of the deflection. This kind of systematics is discussed below.

\begin{description}
\item[Tracking system resolution:] the high threshold for protons in the RICH (18~GV) permitted us to study several features of the tracking system up to high rigidity. The momentum resolution for protons using the RICH information was comparable to the resolution of the tracking system up to 50~GV, and in the region from 19 to 25~GV even superior. The RICH detector was therefore used to cross--check the momentum measurement done by the tracking system. It is important to point out that the CAPRICE98 experiment is till now the only balloon or space experiment capable of cross--checking the rigidity measurement above 5~GV during the flight.

The following procedure was used. The distribution of the Cherenkov angle for protons obtained from flight data in several rigidity bins, which were selected with the tracker system, was compared to the distribution obtained from a simulation. In the simulation an offset parameter in the deflection obtained from the simulated tracking system was varied over a large offset range. 

The simulation program worked as described below: 
\begin{itemize}
  \item from the pre--flight data with the magnet off, the resolution functions of the spectrometer and of the RICH system were obtained by selecting muons;
  \item the pressure of the gas inside the RICH varied during the flight and a distribution of the refractive index of the C$_4$F$_{10}$ gas was obtained from the flight pressure data;
  \item a proton power law spectrum was simulated with a spectral index of -2.7. The rigidity of each simulated proton, R$_{\mathrm{sim}}$, was transformed into deflection, and after being smeared with the tracking resolution function, it was transformed back to rigidity, R$_{\mathrm{sim}}^{\mathrm{s}}$.
  \item The Cherenkov angle for proton was calculated using the simulated rigidity, R$_{\mathrm{sim}}$, and a value of the refractive index was obtained from the distribution of refractive index from the data. Then the Cherenkov angle was smeared according to the experimental resolution function of the RICH.
  \item This calculated Cherenkov angle was plotted in the bin corresponding to the simulated tracking rigidity, R$_{\mathrm{sim}}^{\mathrm{s}}$. 
\end{itemize}

A comparison was made between the real data and the expected ones in each rigidity bin for different offset values. If an offset existed, it was found to be not larger than 7$\times$10$^{-4}$~GV$^{-1}$ with a confidence level of 90\%. 

\begin{figure}[!hbtp]
\includegraphics[width=1.\textwidth]{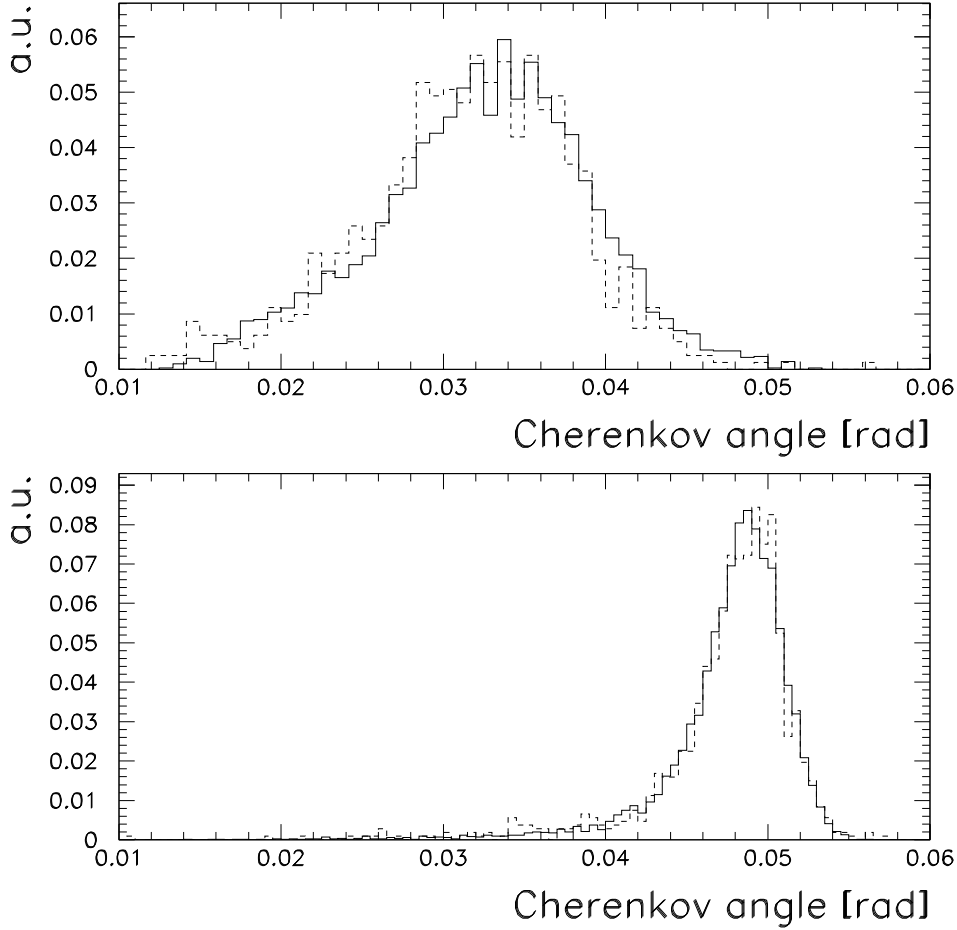}
\caption{The Cherenkov angle distribution in the rigidity bin from 21.4 to 22.9~GV is shown at the top, while at the bottom is shown for a bin from 40.5 to 53.4~GV; here, the dashed lines are for real data and solid lines are for simulated data.}
\label{chere}
\end{figure}

Figure \ref{chere} shows good agreement between the Cherenkov angle distributions in the case of real data (dashed lines) and simulated ones (solid lines). The upper part of this figure is for the rigidity bin from 21.4 to 22.9~GV while the lower part is for the rigidity bin from 40.5 to 53.4~GV. The long tail toward lower angle in the second case is due to the effect of the finite resolution of the spectrometer, indicating the presence of events in that bin with lower rigidity. The effect is more important at high rigidities and is not visible in the lower rigidity bin 21.4--22.9~GV. The good agreement between simulation and real data confirms that the resolution function of spectrometer obtained from the ground data is reliable and is also applicable for the float data.

The simulation was tested also with different spectral indices such as -2.6 and -2.8 and no significant variations were found.
\end{description}

%//////////////////////////////////////////////////////////////////////////////
\subsubsection{Conclusions on systematic uncertainties}
%//////////////////////////////////////////////////////////////////////////////
From the above discussion, and by assuming that the systematic errors are uncorrelated, we estimated that the measurements could have systematic uncertainties that are energy dependent, but less than 10\% below 200~GeV and less than 13\% below 350~GeV.

This estimated uncertainty is not included in the flux values shown in the figures and in the tables.

%##############################################################################
\section{Results}
\label{finres}
%##############################################################################
Given the number of events selected with the proton and helium criteria ($N^{TOA}_{p,He}$), which were corrected for selection efficiencies, transmission function, losses in the payload and in the atmosphere and for atmospheric secondaries, we obtained the fluxes at the top of the atmosphere from the relation

\begin{equation}
Flux(E)_{p,He}=\frac{1}{T_{live}\times G\times \Delta E}\times N^{TOA}_{p,He}(E),
\end{equation}

where $\Delta E$ is the energy bin corrected for ionization losses to the top of the atmosphere and $E$ the kinetic energy in the case of protons or kinetic energy per nucleon for helium nuclei.

\begin{figure}[!hbtp]
\includegraphics[width=1.\textwidth]{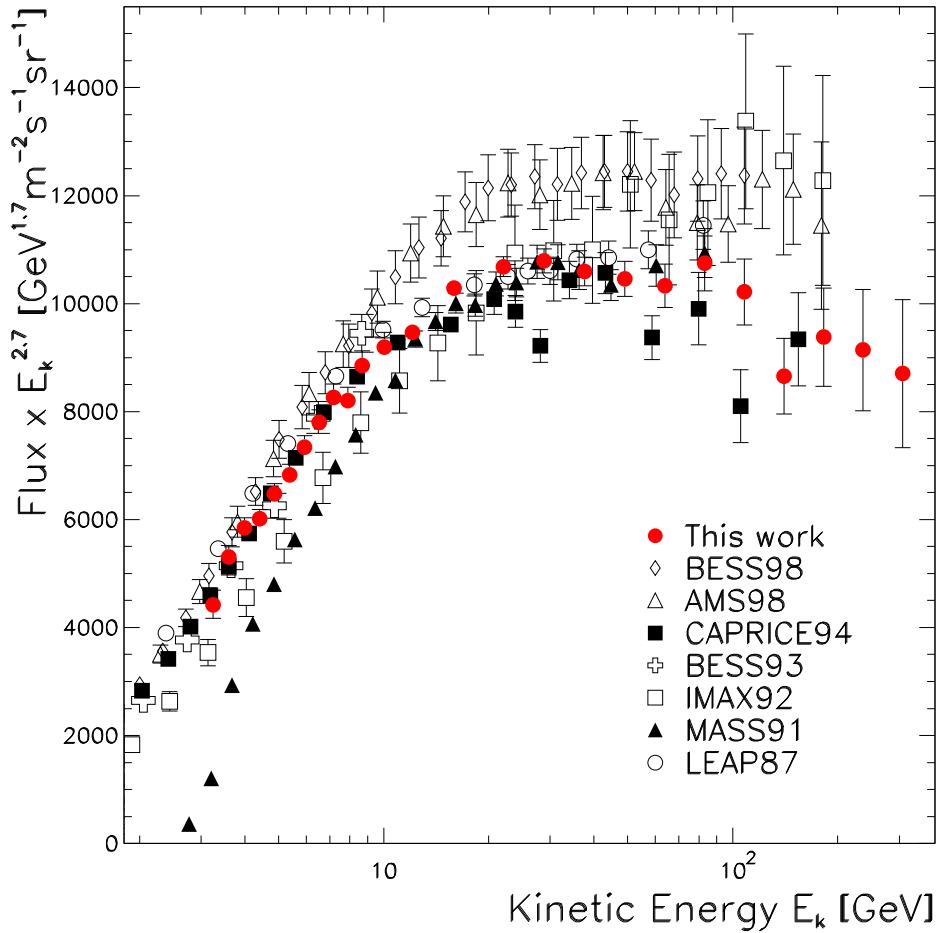}
\caption{The proton energy spectrum at the top of atmosphere detected by CAPRICE98. Results from other recent experiments are also shown: BESS98 \cite{sanuki}, AMS98 \cite{alcaraz}, CAPRICE94 \cite{boezio99b}, BESS93 \cite{bess93}, IMAX92 \cite{menn}, MASS91 \cite{bellotti}, LEAP87 \cite{seo}.}
\label{pflux}
\end{figure}

\begin{figure}[!hbtp]
\includegraphics[width=1.\textwidth]{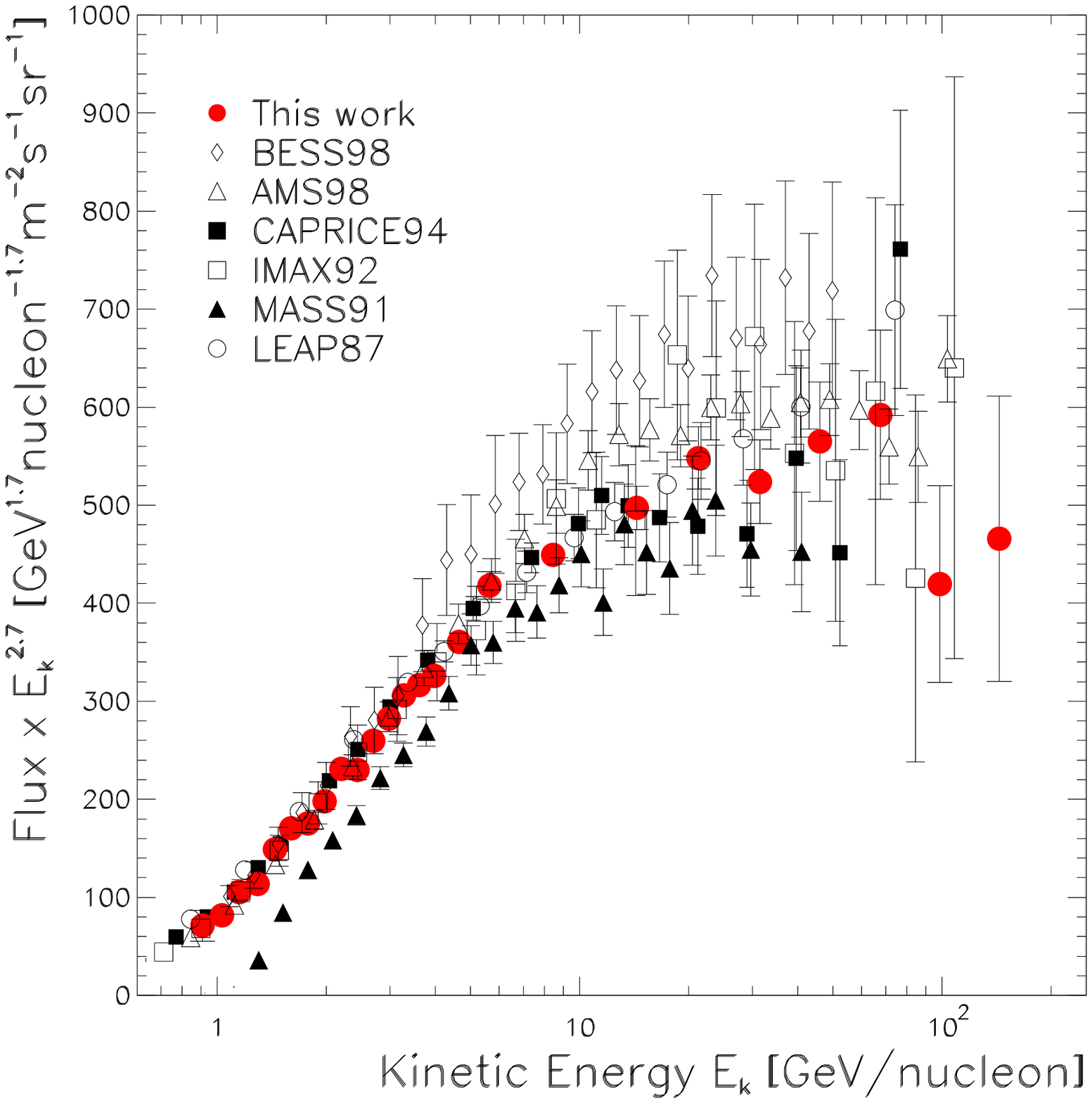}
\caption{The helium energy spectrum at the top of atmosphere detected by CAPRICE98. Results from other recent experiments are also shown: BESS98 \cite{sanuki}, AMS98 \cite{alcaraz}, CAPRICE94 \cite{boezio99b}, IMAX92 \cite{menn}, MASS91 \cite{bellotti}, LEAP87 \cite{seo}.}
\label{heflux}
\end{figure}

The resulting fluxes are given in tables \ref{tabpflu} and \ref{tabheflu}. Figures \ref{pflux} and \ref{heflux} show the energy spectrum of primary cosmic ray protons and helium nuclei respectively along with results from other recent experiments. Systematic errors are also included in the error bars of the BESS98 \cite{sanuki} and AMS \cite{alcaraz} experiments.

% TABLE 4 / TABLE 5

%&&&&&&&&&&&&&&&&&&&&&&&&&&&&&&&&&&&&&&&&&&&&&&&&&&&&&&&&&&&&&&&&&&&&&&&&&&&&&&
\subsection{Proton spectrum at the top of the atmosphere}
%&&&&&&&&&&&&&&&&&&&&&&&&&&&&&&&&&&&&&&&&&&&&&&&&&&&&&&&&&&&&&&&&&&&&&&&&&&&&&&
Solar modulation effect is expected to be very small at rigidities above 20~GV and hence the observed proton spectrum above 20~GeV, which can be fitted by a single power law in kinetic energy, represents the power--law interstellar spectrum. A power law fit to our flux data between 20 and 350~GeV gives
\begin{equation}
J(E)=(1.27\pm 0.09)~\times~10^{4}~E~^{-2.75\pm0.02}~(\mbox{m}^2~\mbox{GeV~sr~s})^{-1}~,
\end{equation}
where $E$ is the kinetic energy; the two variables resulting from the fit are strongly correlated with a correlation coefficient of 0.998.

We notice a nice agreement of the present data with the measurements performed by the WiZard collaboration in previous experiments (MASS91 \cite{bellotti} and CAPRICE94 \cite{boezio99b}. A difference of about 10\% found between CAPRICE94 and CAPRICE98, is considered to be in good agreement, as the statistical and systematic errors, determined by Boezio et al. \cite{boezio99b} were of the order of 10\%. There is also a satisfactory agreement between the~CAPRICE98 results and the other experiments shown in figure \ref{pflux}.

However, certain discrepancies may be found when comparing the present data with the BESS98 and AMS proton data. Both are higher than our results for energies larger than 10~GeV at a level which is barely considered to be consistent with the estimated uncertainties. This occurrence is intriguing, since the agreement between the CAPRICE98 and the BESS98 results, as well as between the CAPRICE94 and BESS98 results, seems to improve at lower energies, but, there is a difference of about 20\% at 100~GeV. Moreover, the BESS93 \cite{bess93} and the BESS94 \cite{bess94} results in the energy range between 1 and 10~GeV are in good agreement with the CAPRICE94 result, which was flown a few days after BESS94. The AMS data \cite{alcaraz} also converge to the CAPRICE98 data below 10~GeV; above this energy the AMS results are about 10--15\% higher. It is interesting to notice that the data published first by AMS \cite{alcaraz1} were lower by about 8\% compared to data published later \cite{alcaraz}, the earlier ones are in good agreement with the CAPRICE98 results, even above 10~GeV, but no comment was made by them regarding the revision of results~in~\cite{alcaraz}.

\begin{figure}[!hbtp]
\includegraphics[width=1.\textwidth]{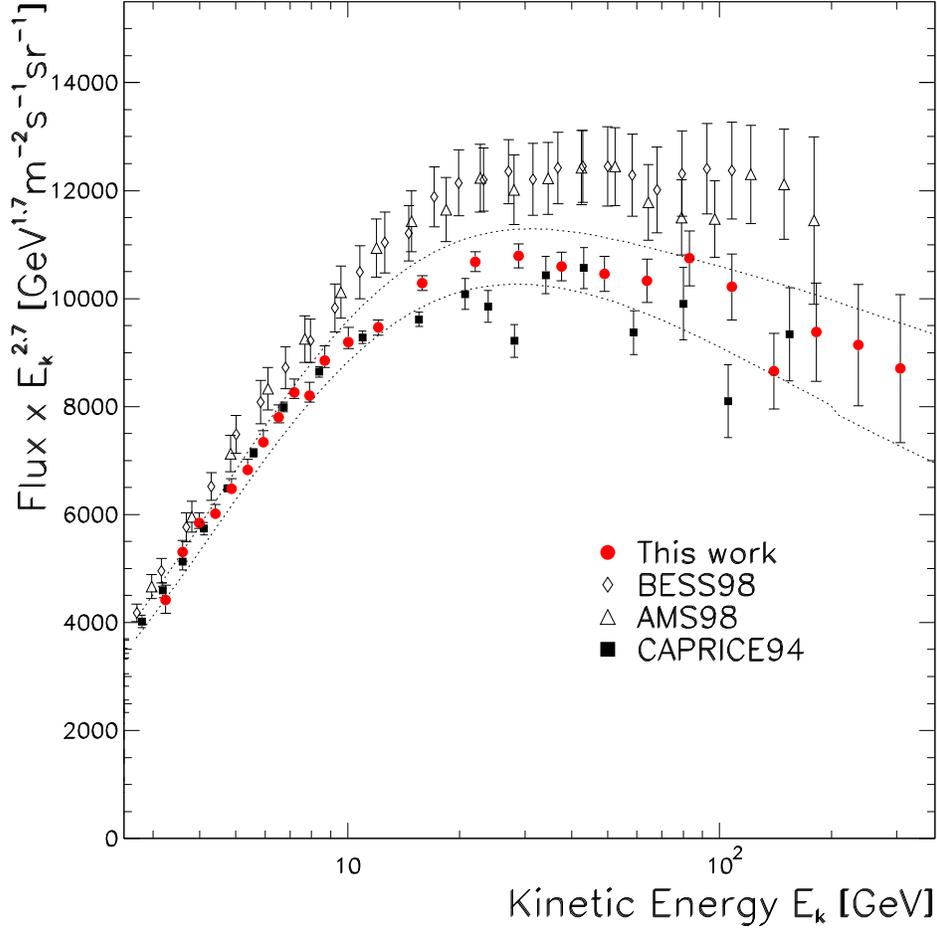}
\caption{The proton energy spectrum at the top of atmosphere detected by CAPRICE98. The upper and lower limit for the estimated systematic errors, described in section \ref{syserr}, are shown as dotted lines; they include an offset in the deflection measurement of 7 $\times$ 10$^{-4}$ GV$^{-1}$. The error bars represent the statistical errors plus the systematics due to efficiencies (section \ref{efficiencies}), see table \ref{tabpflu}. Results from BESS98, AMS98 and CAPRICE94 experiments are also shown.}
\label{statpfl}
\end{figure}
Figure \ref{statpfl} shows the proton energy spectrum measured by CAPRICE98, along with the maximal estimated systematic uncertainties, which are shown by dotted curves. An offset in the deflection of $\pm$ 7 $\times$ 10$^{-4}$ GV$^{-1}$, corresponding to the extreme value of the confidence level derived in section \ref{syserr}, is included in the systematic uncertainties.

Using the equation proposed by Gaisser et al. \cite{gaisser01} for the differential spectrum
\begin{equation}
\label{gaisequ}
J(E)=K \Big( E+b \exp \Big[ -c \sqrt{E} \Big] \Big)^{-\alpha},
\end{equation}
where $E$ is the kinetic energy per nucleon (GeV), $b=2.15$~GeV and $c=0.21$ GeV$^{-0.5}$, the fit of the CAPRICE98 proton spectrum gives $K=1.46\pm0.03$~$\times$~10$^4$ (GeV$^{\alpha-1}$ m$^2$ sr s)$^{-1}$ and $\alpha=2.776\pm0.002$; this result can be compared to the fit of the combined BESS and AMS data that gave $K=1.49\pm0.06$~$\times$~10$^4$~(GeV$^{\alpha-1}$ m$^2$ sr s)$^{-1}$ and $\alpha=2.74\pm0.01$ \cite{gaisser01}. As can be seen there is a good agreement in the value of the constant $K$, while in the case of the spectral index $\alpha$ there is a two sigma difference. The good agreement of the constant is due to  high statistics at the low energy part of the spectrum that has a greater influence on the fit, where the spectra converge, than the high energy part. 

%&&&&&&&&&&&&&&&&&&&&&&&&&&&&&&&&&&&&&&&&&&&&&&&&&&&&&&&&&&&&&&&&&&&&&&&&&&&&&&
\subsection{Helium nuclei spectrum at the top of the atmosphere}
%&&&&&&&&&&&&&&&&&&&&&&&&&&&&&&&&&&&&&&&&&&&&&&&&&&&&&&&&&&&&&&&&&&&&&&&&&&&&&&
The flux data on helium can be fitted by a power law spectrum between 15 and 150~GeV/n. 
%As in the case of protons, the lower bound correspond to a rigidity of about 20~GV and thus 
The fitted spectrum represents the power--law interstellar spectrum in kinetic energy per nucleon and is given by
\begin{equation}
J(E)=(4.8\pm 0.8)~\times~10^{2}~E~^{-2.67\pm0.06}~(\mbox{m}^2~\mbox{GeV~nucleon}^{-1}\mbox{~sr~s})^{-1}~,
\end{equation}
where $E$ is the kinetic energy per nucleon; as in the case of protons, the two variables are strongly correlated with a correlation coefficient of 0.998.

In the case of helium nuclei (figure \ref{heflux}) measurements are in better agreement, even though BESS98 flux is still higher than the other ones. In this case, BESS98 data are higher than AMS data by about 15\%, while the CAPRICE98 spectrum converge to the BESS98 data below 3~GeV/nucleon, but at 50~GeV/nucleon, it is lower by about 20\%.

Using the equation \ref{gaisequ}, with $b=1.50$~GeV and $c=0.30$ GeV$^{-0.5}$, a good agreement with the BESS and AMS fit \footnote{The ``low'' fit in \cite{gaisser01}.} is found in the case of helium nuclei. The CAPRICE98 parameters are $K=6.28\pm0.14$~$\times$~10$^2$~(GeV$^{\alpha-1}$ nucleon$^{1-\alpha}$ m$^2$ sr s)$^{-1}$ and $\alpha=2.753\pm0.014$ while for the BESS/AMS fit the parameters are $K=7.5\pm1.0$~$\times$~10$^2$~(GeV$^{\alpha-1}$ nucleon$^{1-\alpha}$ m$^2$ sr s)$^{-1}$ and $\alpha=2.74\pm0.03$.

%&&&&&&&&&&&&&&&&&&&&&&&&&&&&&&&&&&&&&&&&&&&&&&&&&&&&&&&&&&&&&&&&&&&&&&&&&&&&&&
\subsection{Proton to helium nuclei ratio}
\begin{figure}[!hbtp]
\includegraphics[width=1.\textwidth]{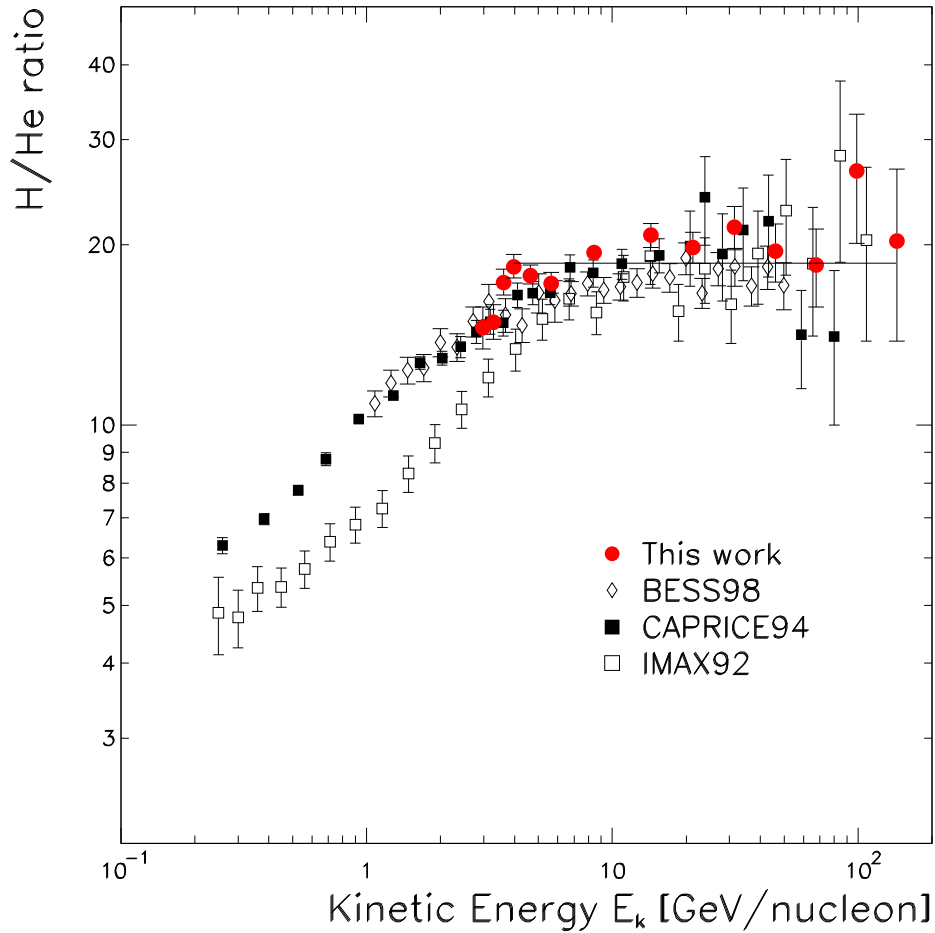}
\caption{Proton to helium nuclei flux ratio at the top of the atmosphere as a function of kinetic energy per nucleon. Results from other recent experiments are also shown: BESS98 \cite{sanuki}, CAPRICE94 \cite{boezio99b}, IMAX92 \cite{menn}.}
\label{hheratio}
\end{figure}

%&&&&&&&&&&&&&&&&&&&&&&&&&&&&&&&&&&&&&&&&&&&&&&&&&&&&&&&&&&&&&&&&&&&&&&&&&&&&&&
Figure \ref{hheratio} shows the proton to helium nuclei ratio at the top of the atmosphere as a function of kinetic energy per nucleon.

The results from the CAPRICE98 experiment show that the proton to helium ratio is nearly constant above 4~GeV~nucleon$^{-1}$ and its value is $18.6 \pm 0.3$, solid line, in agreement with the CAPRICE94 result \cite{boezio99b}. Also shown are data by BESS98 \cite{sanuki} and IMAX92 \cite{menn}. Although there is a large spread in the data points, one finds a general agreement between these measurements. It can be noted from this figure a change in the ratio between proton and helium nuclei around 4~GeV~nucleon$^{-1}$. This transition may have important implications on the propagation of cosmic rays.

%&&&&&&&&&&&&&&&&&&&&&&&&&&&&&&&&&&&&&&&&&&&&&&&&&&&&&&&&&&&&&&&&&&&&&&&&&&&&&&
\subsection{Discussion of results}
%&&&&&&&&&&&&&&&&&&&&&&&&&&&&&&&&&&&&&&&&&&&&&&&&&&&&&&&&&&&&&&&&&&&&&&&&&&&&&&
A possible explanation for the difference between various experiments above\linebreak 10~GeV could be that there are additional uncertainties related to the calibration of the tracking systems. We note that an error in the rigidity of only 3\% produces an error of about 10\% in the high energy tail of the fluxes.

We may point out that the finite resolution of the spectrometer has a different effect from a possible offset in the deflection measurement. For example, in the case of CAPRICE98 the finite resolution causes a deformation of the final flux of the order of 1\% at 100~GeV, while an offset of 1~$\times$10$^{-3}$~GV$^{-1}$ in the deflection measurement would produce a change in the flux of about 10\% at 100~GeV.

The systematic errors taken into account in the BESS98 experiment \cite{sanuki} did not include any possible systematic error in the rigidity measurement and no comments on the alignment of the chambers were made in their paper. The BESS spectrometer was tested at KEK beam with protons and anti--protons at momenta below 2~GeV/c \cite{asaoka}, in February 1999 after the flight. However, from the flight data, it could have been possible only to cross check the rigidity measurement over a limited rigidity range up to a few GV, where the TOF was able to determine the particle velocity.

The calibration of the AMS spectrometer was done before and after the flight using proton and pion beams with momentum from 2 to 14~GeV/c and during the flight the alignment was monitored with an infrared laser system \cite{alcaraz2}. Since it was possible to determine an upper limit to the misalignment of the silicon tracker detector, it should be possible to determine an upper limit for an eventual offset in the deflection measurement. But no systematic error related to the alignment of the silicon plane was considered in the AMS data \cite{alcaraz,alcaraz1}. For example, in the case of CAPRICE97, an offset of 5~$\times$10$^{-4}$~GV$^{-1}$ was measured with a misalignment of 7~$\mu$m of the middle drift chamber \cite{kremer}.

In spite of these discrepancies, we may note that the level of agreement among these recent measurements is within 10--20\%, which is indeed a significant improvement with respect to previous years. This is particularly important in considering the fact that these results, including that of the CAPRICE98, are significantly lower than some of the older measurements (e.g., Webber \cite{webber} not shown in figure \ref{pflux}).

%##############################################################################
\section{Conclusions}
%##############################################################################
The primary proton and helium nuclei results from the latest CAPRICE balloon--borne experiment, performed in 1998, were presented. The excellent performance of this apparatus allowed an accurate measurement of the spectra extended over a large energy range. For the first time it was possible to cross--check the rigidity measurement during the flight at high energy, where the measured spectra are more sensitive to this kind of systematic errors. The CAPRICE98 instrument made it possible to accurately determine efficiencies, rejection power and to estimate systematic errors for each individual detector. This allowed us to measure proton and helium nuclei spectra with an excellent understanding of the performance of the detectors. The results are in good agreement with other recent measurements if the systematic errors are properly taken into account. However, all these results are significantly lower than some of the older measurements.

New space experiments, PAMELA \cite{pamela} in 2003 and AMS2 \cite{ams2} later on the International Space Station, will be able to perform the same measurements with better statistics than was done in the past, but the CAPRICE98 instrument will remain a unique detector that joined together a precise superconducting magnet spectrometer and a high threshold gas RICH detector, an excellent apparatus to study\linebreak cosmic--rays.

\ack
This work was supported by NASA Grant NADW--110, the Istituto Nazionale di Fisica Nucleare, Italy, the Agenzia Spaziale Italiana, DARA and DFG in Germany, the Swedish National Space Board and the Knut and Alice Wallenberg foundation. The Swedish--French group thanks the EC SCIENCE programme for support. P.H. was supported by the Swedish Foundation for International Cooperation in Research and Higher Education; E.M. was supported by the Foundation BLANCEFLOR Boncompagni--Ludovisi, n\'ee Bildt. We wish to thank the National Scientific Balloon Facility and the NSBF launch crew that served in Fort Sumner. We would also like to acknowledge the essential support given by the Gas Work Group (EST/SM/SF) and the Thin Films \& Section (EP/TA1/TF) at CERN, the LEPSI and CRN--Strasbourg and the technical staff of NMSU and of INFN.

%##############################################################################

%##############################################################################
%     TABLES 
%##############################################################################

% TABLE 1
\newpage
\hoffset=-1.5cm
\begin{table}[!hbt]
\begin{tabular}{ccccc}
\hline
\hline
Rigidity at & Observed Number of & Extrapolated number & Atmospheric & Extrapolated number of\\
spectrometer $[$GV$]$ & events at spectrometer & at top of payload & correction &primary events at TOA$^{\mbox{a}}$ \\
\hline
3.88 \mbox{--} 4.23 & 9106 & 12779 & 1570 & 13553 \\
4.23 \mbox{--} 4.61 & 15344 & 21542 & 2000 & 22851 \\
4.61 \mbox{--} 5.02 & 17648 & 24786 & 2112 & 26297 \\
5.02 \mbox{--} 5.47 & 16288 & 22885 & 1920 & 24285 \\
5.47 \mbox{--} 5.96 & 14962 & 21031 & 1741 & 22321 \\
5.96 \mbox{--} 6.49 & 13220 & 18590 & 1533 & 19734 \\
6.49 \mbox{--} 7.08 & 11923 & 16774 & 1372 & 17809 \\
7.08 \mbox{--} 7.71 & 10660 & 15004 & 1220 & 15933 \\
7.71 \mbox{--} 8.40 & 9487 & 13359 & 1082 & 14189 \\
8.40 \mbox{--} 9.15 & 7965 & 11221 & 916 & 11920 \\
9.15 \mbox{--} 10.0 & 7263 & 10238 & 827 & 10877 \\
10.0 \mbox{--} 12.0 & 12545 & 17698 & 1438 & 18808 \\
12.0 \mbox{--} 14.0 & 7748 & 10943 & 893 & 11634 \\
14.0 \mbox{--} 20.2 & 12653 & 17548 & 1438 & 18664 \\
20.2 \mbox{--} 26.1 & 5048 & 7269 & 597 & 7736 \\  
\end{tabular}
\end{table} 
\newpage
\begin{table}[!hbt]
\begin{tabular}{ccccc}
\mbox{\hspace{17ex}} & \mbox{\hspace{20ex}}&\mbox{\hspace{19ex}} &\mbox{\hspace{11ex}} & \mbox{\hspace{20ex}}\\
26.1 \mbox{--} 33.9 & 3150 & 4627 & 383 & 4926 \\  
33.9 \mbox{--} 44.0 & 1930 & 2868 & 240 & 3056 \\  
44.0 \mbox{--} 57.0 & 1197 & 1792 & 152 & 1910 \\  
57.0 \mbox{--} 73.9 & 748 & 1126 & 96 & 1200 \\   
73.9 \mbox{--} 95.8 & 493 & 746 & 64 & 795 \\
95.8 \mbox{--} 124 & 298 & 453 & 40 & 484 \\
124 \mbox{--} 161 & 161 & 246 & 23 & 263 \\ 
161 \mbox{--} 209 & 111 & 171 & 15 & 182 \\ 
209 \mbox{--} 270 & 69 & 107 & 10 & 114 \\ 
270 \mbox{--} 351 & 42 & 65 & 6 & 70 \\ 
\hline
\hline
\end{tabular}
\vspace{0.5cm}
\caption{Summary of proton results. $^{\mbox{a}}$ Top of the atmosphere. The extrapolated number of events at the top of payload includes the correction for detector efficiencies.}
\label{tabcorr}
\end{table}

%&&&&&&&&&&&&&&&&&&&&&&&&&&&&&&&&&&&&&&&&&&&&&&&&&&&&&&&&&&&&&&&&&&&&&&&&&&&&&&

% TABLE 2
\newpage
\begin{table}[!hbt]
\begin{tabular}{ccccc}
\hline
\hline
Rigidity at & Observed Number of & Extrapolated number & Atmospheric & Extrapolated number of\\
spectrometer $[$GV$]$ & events at spectrometer & at top of payload & correction &primary events at TOA \\
\hline
3.00 \mbox{--} 3.25 & 77 & 114 & 11 & 124 \\ 
3.25 \mbox{--} 3.52 & 196 & 291 & 27 & 317 \\ 
3.52 \mbox{--} 3.81 & 501 & 745 & 70 & 813 \\ 
3.81 \mbox{--} 4.13 & 988 & 1472 & 139 & 1606 \\ 
4.13 \mbox{--} 4.48 & 1812 & 2705 & 256 & 2952 \\ 
4.48 \mbox{--} 4.85 & 2225 & 3328 & 314 & 3632 \\ 
4.85 \mbox{--} 5.26 & 2083 & 3123 & 295 & 3408 \\ 
5.26 \mbox{--} 5.69 & 1988 & 2987 & 282 & 3260 \\ 
5.69 \mbox{--} 6.17 & 1921 & 2894 & 273 & 3158 \\ 
6.17 \mbox{--} 6.68 & 1581 & 2388 & 226 & 2605 \\ 
6.68 \mbox{--} 7.24 & 1490 & 2256 & 213 & 2461 \\ 
7.24 \mbox{--} 7.84 & 1346 & 2043 & 193 & 2229 \\ 
7.84 \mbox{--} 8.50 & 1210 & 1841 & 174 & 2009 \\ 
\end{tabular}
\end{table} 
\newpage
\begin{table}[!hbt]
\begin{tabular}{ccccc}
\mbox{\hspace{17ex}} & \mbox{\hspace{20ex}}&\mbox{\hspace{19ex}} &\mbox{\hspace{11ex}} & \mbox{\hspace{20ex}}\\
8.50 \mbox{--} 9.21 & 1050 & 1602 & 151 & 1748 \\ 
9.21  \mbox{--} 9.97 & 907 & 1388 & 131 & 1514 \\ 
9.97 \mbox{--} 12.0 & 1739 & 2672 & 252 & 2916 \\ 
12.0 \mbox{--} 14.0 & 1175 & 1815 & 172 & 1981 \\ 
14.0 \mbox{--} 25.4 & 2410 & 3761 & 355 & 4105 \\ 
25.4 \mbox{--} 37.0 & 637 & 1008 & 95 & 1099 \\ 
37.0 \mbox{--} 53.8 & 350 & 558 & 53 & 609 \\ 
53.8 \mbox{--} 78.3 & 170 & 272 & 26 & 297 \\ 
78.3 \mbox{--} 114 & 94 & 151 & 14 & 165 \\ 
114 \mbox{--} 166 & 51 & 82 & 8 & 90 \\ 
166 \mbox{--} 241 & 19 & 31 & 3 & 33 \\ 
241 \mbox{--} 351 & 11 & 18 & 2 & 19 \\
\hline
\hline
\end{tabular}
\vspace{0.5cm}
\caption{Summary of helium nuclei results. The extrapolated number of events at the top of payload includes the correction for detector efficiencies.}
\label{tabcorrhe}
\end{table} 

%&&&&&&&&&&&&&&&&&&&&&&&&&&&&&&&&&&&&&&&&&&&&&&&&&&&&&&&&&&&&&&&&&&&&&&&&&&&&&&

% TABLE 3
\newpage
\hoffset=0cm
\begin{table}[!hbt]
\begin{tabular}{lcc}
\hline
\hline
Source & Rigidity range $[$GV$]$ & Estimate $[$\%$]$ \\
\hline
Z$=$1 particle contamination in proton sample & 2 -- 350 & $<0.4$ \\
\hline
Z$>$1 particle contamination in proton sample & 2 -- 350 & $<0.2$ \\
\hline
Z$\ne$2 particle contamination in helium sample & 2 -- 350 & $<0.1$ \\
\hline
Trigger efficiency & 2 -- 350 & $\le 2$ \\
\hline
Geometrical factor & 2 -- 350 & $\le 2$ \\
\hline
Atmospheric secondaries & 2 -- 350 & $\le 1$ \\ 
Uncertainty on residual atmosphere & 2 -- 350 & $\le 1.5$ \\
\hline
Losses in the atmosphere and payload & 2 -- 350 & $\le 2$ \\
Uncertainty on atmospheric depths & 2 -- 350 & $\le 1$ \\
\hline
Spectrometer resolution & 2 -- 200 & $\le 2$ \\
                        & 200 -- 350 & $\le 5$ \\
\hline
Tracking system & 2 -- 100 & $\le 6$ \\
                & 100 -- 350 & $\le 13$ \\ 
\hline
\hline
\end{tabular}
\vspace{0.5cm}
\caption{Estimate of systematic errors on the fluxes and of contaminations.}
\label{ester}
\end{table} 

%&&&&&&&&&&&&&&&&&&&&&&&&&&&&&&&&&&&&&&&&&&&&&&&&&&&&&&&&&&&&&&&&&&&&&&&&&&&&&&

% TABLE 4
\newpage
\begin{table}[!hbt]
\begin{center}
\begin{tabular}{ccccr@{ $\times$ }l}
\hline
\hline
\multicolumn{3}{c}{Kinetic Energy}  & Mean Kinetic Energy &\multicolumn{2}{c}{Proton Flux at TOA} \\
\multicolumn{3}{c}{at TOA $[$GeV$]$} & at TOA $[$GeV$]$ & \multicolumn{2}{c}{$[$(m\sq sr s GeV)$^{-1}]$}\\
\hline
3.08 &--& 3.42 \rule{0pt}{3ex} & 3.24 & $\Big($ 1.85 $^{+\mbox{ 0.11 }}_{- \mbox{ 0.10 }} \Big)$& 10$^{2}$ \\ 
3.42 &--& 3.78 \rule{0pt}{3ex} & 3.60 & $\Big($ 1.67 $^{+\mbox{ 0.07 }}_{- \mbox{ 0.05 }} \Big)$& 10$^{2}$ \\ 
3.78 &--& 4.19 \rule{0pt}{3ex} & 3.99 & $\Big($ 1.39 $^{+\mbox{ 0.04 }}_{- \mbox{ 0.02 }} \Big)$& 10$^{2}$ \\ 
4.19 &--& 4.64 \rule{0pt}{3ex} & 4.41 & $\Big($ 1.09 $^{+\mbox{ 0.03 }}_{- \mbox{ 0.01 }} \Big)$& 10$^{2}$ \\ 
4.64 &--& 5.12 \rule{0pt}{3ex} & 4.87 & $\Big($ 9.02 $^{+\mbox{ 0.26 }}_{- \mbox{ 0.11 }} \Big)$& 10$^{1}$ \\ 
5.12 &--& 5.65 \rule{0pt}{3ex} & 5.38 & $\Big($ 7.27 $^{+\mbox{ 0.21 }}_{- \mbox{ 0.09 }} \Big)$& 10$^{1}$ \\ 
5.65 &--& 6.22 \rule{0pt}{3ex} & 5.93 & $\Big($ 6.00 $^{+\mbox{ 0.17 }}_{- \mbox{ 0.08 }} \Big)$& 10$^{1}$ \\ 
6.22 &--& 6.85 \rule{0pt}{3ex} & 6.53 & $\Big($ 4.92 $^{+\mbox{ 0.14 }}_{- \mbox{ 0.07 }} \Big)$& 10$^{1}$ \\ 
6.85 &--& 7.54 \rule{0pt}{3ex} & 7.19 & $\Big($ 4.02 $^{+\mbox{ 0.12 }}_{- \mbox{ 0.06 }} \Big)$& 10$^{1}$ \\ 
7.54 &--& 8.29 \rule{0pt}{3ex} & 7.90 & $\Big($ 3.09 $^{+\mbox{ 0.09 }}_{- \mbox{ 0.04 }} \Big)$& 10$^{1}$ \\ 
8.29 &--& 9.10 \rule{0pt}{3ex} & 8.68 & $\Big($ 2.59 $^{+\mbox{ 0.08 }}_{- \mbox{ 0.04 }} \Big)$& 10$^{1}$ \\ 
9.10 &--& 11.1 \rule{0pt}{3ex} & 10.0 & $\Big($ 1.81 $^{+\mbox{ 0.05 }}_{- \mbox{ 0.02 }} \Big)$& 10$^{1}$ \\ 
\end{tabular}
\end{center}
\end{table} 
\newpage
\begin{table}[!hbt]
\begin{center}
\begin{tabular}{ccccr@{ $\times$ }l}
\multicolumn{3}{c}{\mbox{\hspace{14ex}}}& \mbox{\hspace{19ex}} & \multicolumn{2}{c}{\mbox{\hspace{18ex}}}\\
11.1 &--& 13.1 \rule{0pt}{3ex} & 12.1 & ( 1.13 $\pm$ 0.02 ) &  10$^{1}$ \\ 
13.1 &--& 19.3 \rule{0pt}{3ex} & 15.9 & ( 5.89 $\pm$ 0.08 ) &  10$^{0}$ \\ 
19.3 &--&  25.2 \rule{0pt}{3ex} & 22.1 & ( 2.52 $\pm$ 0.04 ) &  10$^{0}$ \\ 
25.2 &--& 33.0 \rule{0pt}{3ex} & 28.8 & ( 1.24 $\pm$ 0.02 ) &  10$^{0}$ \\ 
33.0 &--& 43.0 \rule{0pt}{3ex} & 37.6 & ( 5.91 $\pm$ 0.15 ) &  10$^{-1}$ \\ 
43.0 &--& 56.1 \rule{0pt}{3ex} & 49.1 & ( 2.85 $\pm$ 0.09 ) &  10$^{-1}$ \\ 
56.1 &--& 73.0 \rule{0pt}{3ex} & 63.9 & ( 1.38 $\pm$ 0.05 ) &  10$^{-1}$ \\ 
73.0 &--& 94.9 \rule{0pt}{3ex} & 83.1 & ( 7.1 $\pm$ 0.3 ) &  10$^{-2}$ \\ 
94.9 &--& 123 \rule{0pt}{3ex} & 108 & ( 3.31 $\pm$ 0.20 ) &  10$^{-2}$ \\ 
123 &--& 160 \rule{0pt}{3ex} & 140 & ( 1.38 $\pm$ 0.11 ) &  10$^{-2}$ \\ 
160 &--& 208 \rule{0pt}{3ex} & 182 & ( 7.4 $\pm$ 0.7 ) &  10$^{-3}$ \\ 
208 &--& 270 \rule{0pt}{3ex} & 236 & ( 3.6 $\pm$ 0.4 ) &  10$^{-3}$ \\ 
270 &--& 350 \rule{0pt}{3ex} & 307 & ( 1.7 $\pm$ 0.3 ) &  10$^{-3}$ \\ 
\hline
\hline
\end{tabular}
\vspace{0.5cm}
\caption{Measured proton flux at the top of the atmosphere. Statistical and efficiency--related systematic errors quadratically summed are reported.}
\label{tabpflu}
\end{center}
\end{table} 

%&&&&&&&&&&&&&&&&&&&&&&&&&&&&&&&&&&&&&&&&&&&&&&&&&&&&&&&&&&&&&&&&&&&&&&&&&&&&&&

% TABLE 5
\newpage
\begin{table}[!hbt]
\begin{center}
\begin{tabular}{r@{\hspace{0.3cm}--\hspace{0.3cm}}lcr@{ $\times$ }l}
\hline
\hline
\multicolumn{2}{c}{Kinetic Energy at}  & Mean Kinetic Energy &\multicolumn{2}{c}{Helium nuclei Flux at TOA} \\
\multicolumn{2}{c}{TOA $[$GeV nucleon$^{-1}]$} & at TOA $[$GeV nucleon$^{-1}]$ & \multicolumn{2}{c}{$[$(m\sq sr s GeV nucleon$^{-1}$)$^{-1}]$}\\
\hline
0.85 & 0.96 & 0.91 & ( 9.1 $\pm$ 2.0 ) & 10$^{1}$\\ 
0.96 & 1.08 & 1.03 & ( 7.6 $\pm$ 1.1 ) & 10$^{1}$\\ 
1.08 & 1.21 & 1.15 & ( 7.2 $\pm$ 0.7 ) & 10$^{1}$\\ 
1.21 & 1.35 & 1.29 & ( 5.7 $\pm$ 0.4 ) & 10$^{1}$\\ 
1.35 & 1.51 & 1.44 & ( 5.6 $\pm$ 0.3 ) & 10$^{1}$\\ 
1.51 & 1.68 & 1.60 & ( 4.78 $\pm$ 0.16 ) & 10$^{1}$\\ 
1.68 & 1.87 & 1.78 & ( 3.69 $\pm$ 0.10 ) & 10$^{1}$\\ 
1.87 & 2.08 & 1.98 & ( 3.13 $\pm$ 0.09 ) & 10$^{1}$\\ 
2.08 & 2.30 & 2.20 & ( 2.75 $\pm$ 0.08 ) & 10$^{1}$\\ 
2.30 & 2.55 & 2.44 & ( 2.07 $\pm$ 0.06 ) & 10$^{1}$\\ 
2.55 & 2.82 & 2.69 & ( 1.79 $\pm$ 0.05 ) & 10$^{1}$\\ 
2.82 & 3.11 & 2.97 & ( 1.49 $\pm$ 0.05 ) & 10$^{1}$\\ 
3.11 & 3.43 & 3.28 & ( 1.24 $\pm$ 0.04 ) & 10$^{1}$\\ 
\end{tabular}
\end{center}
\end{table} 
\newpage
\begin{table}[!hbt]
\begin{center}
\begin{tabular}{r@{\hspace{0.3cm}--\hspace{0.3cm}}lcr@{ $\times$ }l}
\multicolumn{2}{c}{\mbox{\hspace{20ex}}}& \mbox{\hspace{22ex}} & \multicolumn{2}{c}{\mbox{\hspace{22ex}}}\\
3.43 & 3.78 & 3.61 & ( 9.9 $\pm$ 0.3 ) & 10$^{0}$\\ 
3.78 & 4.16 & 3.97 & ( 7.9 $\pm$ 0.3 ) & 10$^{0}$\\ 
4.16 & 5.15 & 4.64 & ( 5.72 $\pm$ 0.17 ) & 10$^{0}$\\ 
5.15 & 6.14 & 5.63 & ( 3.93 $\pm$ 0.14 ) & 10$^{0}$\\ 
6.14 & 11.8 & 8.44 & ( 1.42 $\pm$ 0.04 ) & 10$^{0}$\\ 
11.8 & 17.6 & 14.4 & ( 3.73 $\pm$ 0.17 ) & 10$^{-1}$\\ 
17.6 & 26.0 & 21.3 & ( 1.42 $\pm$ 0.08 ) & 10$^{-1}$\\ 
26.0 & 38.2 & 31.4 & ( 4.8 $\pm$ 0.4 ) & 10$^{-2}$\\ 
38.2 & 56.0 & 46.1 & ( 1.82 $\pm$ 0.20 ) & 10$^{-2}$\\ 
56.0 & 81.9 & 67.5 & ( 6.8 $\pm$ 1.0 ) & 10$^{-3}$\\ 
81.9 & 120 & 99.6 & ( 1.7 $\pm$ 0.4 ) & 10$^{-3}$\\ 
120 & 174 & 144 & ( 6.9 $\pm$ 2.2 ) & 10$^{-4}$\\ 
\hline
\hline
\end{tabular}
\vspace{0.5cm}
\caption{Measured helium nuclei flux at the top of the atmosphere. Statistical and efficiency--related systematic errors quadratically summed are reported.}
\label{tabheflu}
\end{center}
\end{table} 

%&&&&&&&&&&&&&&&&&&&&&&&&&&&&&&&&&&&&&&&&&&&&&&&&&&&&&&&&&&&&&&&&&&&&&&&&&&&&&&

\end{document}